\documentclass[12pt, a4, twoside]{report}
\usepackage{sectsty}
\usepackage{times}
\usepackage{geometry}
\usepackage[utf8]{inputenc}
\usepackage{minitoc}
\usepackage{newunicodechar}
\newunicodechar{′}{\ensuremath{'}}
\usepackage{setspace}
\usepackage[style=apa, backend=biber]{biblatex}
\usepackage{lipsum}
\usepackage{etoolbox}
\usepackage{subcaption}
\usepackage{appendix}
\usepackage{url} 
\usepackage{amsmath}
\usepackage{multirow}
\usepackage[skip=20pt, indent=0pt]{parskip} 
\usepackage{graphicx}
\usepackage{glossaries}
\makeglossaries
\newacronym{NASA}{NASA}{National Aeronautics and Space Administration}
\newacronym{ARCSSTE-E}{ARCSSTE-E}{African Regional Center for Space Science and Technology Education in English}
\newacronym{GHGs}{GHGs}{Greeenhouse Gases}
\newacronym{IPCC}{IPCC}{Intergovernmental Panel on Climate Change}
\newacronym{GSAT}{GSAT}{Global Surface Air Temperature}
\newacronym{POWER}{POWER}{Prediction Of Worldwide Energy Resource}
\newacronym{WCRP}{WCRP}{World Climate Research Programme}
\newacronym{GEWEX}{GEWEX}{Global Energy and Water Cycle Experiment}
\newacronym{SRB}{SRB}{Surface Radiation Budget}
\newacronym{GMAO}{GMAO}{Global Modeling and Assimilation Office}
\newacronym{GEOS}{GEOS}{Goddard Earth Observing System}
\newacronym{fvGCM}{fvGCM}{finite-volume General Circulation Model}
\newacronym{GSFC}{GSFC}{NASA Goddard Space Flight Center}
\newacronym{ML}{ML}{Machine Learning}
\newacronym{PSAS}{PSAS}{Physical-space Statistical Analysis System }
\newacronym{XGBoost}{XGBoost}{eXtreme Gradient Boosting}
\newacronym{GMET}{GMET}{Ghana Meteorological Agency}
\newacronym{RCMs}{RCMs}{Regional Climate Models}
\newacronym{GCMs}{GCMs}{Global Climate Models}
\newacronym{RCP}{RCP}{Representative Concentration Pathway}
\newacronym{LRT}{LRT}{Linear Regression Trend}
\newacronym{LST}{LST}{Land Surface Temperature}
\newacronym{LaRC}{LaRC}{Langley Research Center}
\newacronym{T2MDEW}{T2MDEW}{Dew/Frost Point at 2$m$}
\newacronym{T2MWET}{T2MWET}{Wet Bulb Temperature at 2$m$}
\newacronym{TS}{TS}{Earth Skin Temperature}
\newacronym{T2M-Range}{T2M-Range}{Temperature at 2$m$ Range}
\newacronym{QV2M}{QV2M}{Specific Humidity at 2$m$}
\newacronym{RH2M}{RH2M}{Relative Humidity at 2$m$}
\newacronym{PRECTOTCORR}{PRECTOTORR} {Precipitation Corrected}
\newacronym{PS}{PS}{Surface Pressure}
\newacronym{WS10M}{WS10M}{Wind Speed at 10$m$}
\newacronym{WS10M-Range}{W210M-Range}{Wind Speed at 10$m$ Range}
\newacronym{WD10M}{WD10M}{Wind Direction at 10$m$}
\newacronym{WS50M}{WS50M}{Wind Speed at 50$m$}
\newacronym{GEE}{GEE}{Google Earth Engine}
\newacronym{ANOVA}{ANOVA}{Analysis of Variance}
\newacronym{SC}{SC}{Single Channel}
\newacronym{AJAX}{AJAX}{Asynchronous JavaScript AndXML}
\newacronym{GBM}{GBM}{Gradient Boosting Machine}
\newacronym{RMSE}{RMSE}{Root Mean Squared Error}
\newacronym{ESRI}{ESRI}{Environmental Systems Research Institute}

\graphicspath{ {./images/} }

\usepackage{biblatex}
\defbibheading{Reference List}{\chapter*{Reference List}}
\begin{document}


\pagenumbering{roman} 

\allsectionsfont{\normalsize}

\begin{titlepage}
\begin{center}

\vspace{3em}

{\textbf{\parbox{\linewidth}{\centering Global Warming In Ghana's Major Cities Based On Statistical Analysis Of NASA's POWER Over 3-Decades}}}

\vspace{1em}





 \vspace{2em}


\vspace{0.5em}

Author \\
Joshua Attih \\ 

\vspace{1em}



\end{center}
\end{titlepage}


\cleardoublepage
\addcontentsline{toc}{chapter}{Abstract}
The effects of high temperatures due to global warming in various parts of the world has raised great concerns. This study, therefore, investigated the long-term temperature trends in four major cities representing different climatic zones in Ghana. Using data from NASA’s Prediction of Worldwide Energy Resource (POWER) project, comprehensive statistical analyses were conducted to assess local climate warming and its implications for the region. The study employed various techniques, including linear regression trend analysis and machine learning via the eXtreme Gradient Boosting (XGBoost) approach to train, test and predict temperature variations. Furthermore, to visualize temperature patterns, Land Surface Temperature (LST) profile maps were generated from the RSLab platform, which provided enhanced accuracy of 1.52°C compared to traditional methods. The results, revealed local warming trends in Accra, Kumasi, and Kete-Krachi. Accra, being heavily industrialized, experienced the highest temperature variability among the cities. Demographic factors were not significant in explaining temperature trends in these cities. The XGBoost model demonstrated good performance, as evidenced by low Root Mean Square Error (RMSE) scores, indicating its effectiveness in capturing temperature patterns over a span of several decades. Based on mid-2023 projections, Wa unexpectedly had the highest mean temperature among the four cities. The estimated mean temperatures for Accra, Kumasi, Kete-Krachi, and Wa in the mid-year 2023 are 27.86°C, 27.15°C, 29.39°C, and 30.76°C, respectively. These findings have contributed to a better understanding of local climate warming in Ghana’s major cities within a short time projection. The XGBoost model with the predictions made will offer valuable insights for policymakers and local communities, enabling the formulation of effective climate change adaptation and mitigation strategies for sustainable cities.


\cleardoublepage
\phantomsection
\addcontentsline{toc}{chapter}{Table of Contents}
\cleardoublepage
\cleardoublepage
\cleardoublepage
\phantomsection
\addcontentsline{toc}{chapter}{List of Figures}

\cleardoublepage
\phantomsection
\addcontentsline{toc}{chapter}{List of Tables}

\cleardoublepage
\phantomsection
\addcontentsline{toc}{chapter}{List of Abbreviations}
\printglossary[type=\acronymtype,title=List of Abbreviations,nonumberlist]

\addcontentsline{toc}{chapter}{List of Abbreviations}

\cleardoublepage 


\newpage
\pagenumbering{arabic} 
\pretocmd{\input}{\doublespacing}{}{}


\section{Introduction to the Study}
Dating back to the pre-industrial ages, Jean-Baptiste Joseph Fourier suggested 
that the atmosphere of the Earth serves as an envelope of \gls{GHGs} with a temperature that is increasing \parencite{fourier1824remarques}. Today it is known from several sources including the \gls{IPCC} that the warming of the Earth's climate system is unequivocal. The IPCC, since its establishment in the 1990s, has released a series of Assessment Reports, Special Reports, Technical Papers, Methodology Reports, and other products \parencite{ipcc2021technical}.

Climate according to \textcite{sorokhtin2007global} is a set of long-term patterns of changes in the state of the atmosphere with the possible atmospheric state parameters such as pressure, humidity and others. This study focuses on temperature. All of the atmospheric state parameters are directly or indirectly being influenced by man. Usually, global warming which results from climate change is quantified through the average increase in the global surface temperature of the Earth \parencite{shaftel2021overview}. This variable is particularly  important for climate policies following the framing of the Paris Agreement in terms of ambitions to stay below  specific levels of the \gls{GSAT} relative to pre-industrial climate. The prospects  of this have, therefore, been constrained by the stabilization of Carbon dioxide at concentrations around 450,  650, and 1000 ppm which governs the seasonal cycle of temperature within the atmosphere \parencite{lindvall1999models}. Two seasonal cycles can be considered in this  case, (1) a component due to direct solar absorption in the atmosphere; and (2) a component due to the flux of energy from the surface to the atmosphere. 

The large seasonal variations in shortwave radiation at the top of the atmosphere are primarily balanced by an energy flux into the ocean \parencite{dines1917heat,donohoe2013seasonal}. But with the widespread Carbon-inequality across the world, this balance is very negligible hence putting developing countries at risk of global warming. As a developing country located in  West Africa which is highly affected by average annual temperatures; $22^{\circ}C - 32^{\circ}C$ \parencite{kabo2016multiyear} and due to her over-dependence on rain-fed agriculture, compounded by factors of widespread poverty and weak capacity \parencite{asante2014climate, worldbank2014}, Ghana is vulnerable to the adverse effects of high temperature rises in the advent of climate change \parencite{kolio2014durability, saha2014urban}, making it crucial to understand and monitor temperature trends in its major cities. 

The development of  statistical approaches which includes \gls{LRT}, hypothesis tests, and \gls{ML} approaches, focuses this research on examining the phenomenon of global warming in Ghana's major cities through data provided by the \gls{NASA}'s \gls{POWER}  project. The results of which would be compared to Landsat's data to produce \gls{LST} maps of the cities under our study.

\section{Background of the Study}
The POWER project derives its data primarily from  NASA’s \gls{WCRP} / \gls{GEWEX}, \gls{SRB} project (Version 2.9) and the \gls{GMAO} \gls{GEOS} assimilation model (V 4.0.3) V \parencite{zhang2007global}, - whose original version came from a \gls{fvGCM} which are valuable for the investigation of various climate phenomena, such as El Niño events, monsoons, and climate variability on different time scales. fvGCM has a \gls{PSAS}. This new system with its Physics was developed at the Data Assimilation Office, \gls{GSFC} Kiehl et al. (1985, 1998). NASA's POWER project offers a  valuable resource for studying climate-related variables, including surface air temperature, solar radiation, and other meteorological parameters, across different regions of the globe. 

By leveraging the comprehensive and reliable data provided by POWER, this research aims to assess the long-term temperature trends in Ghana's major cities and analyze the extent of global warming within these urban environments. The primary objective of this study is to conduct a thorough statistical analysis of the air temperature data collected by NASA's  POWER project over a span of three decades. By utilizing advanced statistical techniques, such as regression trend analysis, and time series modeling \textcite{stocker2013climate} we will iteratively train and evaluate these data via \gls{XGBoost} an ML approach using an ensemble of weak decision trees, where each new tree is built to correct the errors made by the previous trees \parencite{chen2016xgboost, grinsztajn2022tree}. These findings will identify and quantify the long-term temperature trends in Ghana's major cities which hold significant implications for both policymakers and local communities in the various climatic zones in Ghana hence formulating effective climate change adaptation and mitigation strategies. This can also inform urban planning and infrastructure development, allowing for more resilient and sustainable cities that can withstand the challenges posed by climate change.

\section{Problem Statement}

Global warming is an urgent issue. Due to this, many great sustainable solutions have been designed to mitigate its impacts. In Africa, most studies have applied in-situ data in solving the problem with the POWER data remaining highly unused. Outside the continent \textcite{marzouk2021assessment, aboelkhair2019assessment, westberg2013analysis} have either compare their data with POWER or used POWER as a supplement to their studies. We deem it necessary to put in the effort to validate some of the finding done in Africa specifically in Ghana on the same issue using a worldwide database like POWER. This study therefore analyses temperature Variations in Ghana with the NASA's POWER project, hence Local Warming in its Major Cities located in different climatic zones. Accra is located in the Coastal Zone, Kumasi is located in the Forest Zone, Kete-Krachi is located in the Transition Zone and Wa is located in the Savannah Zone.

\section{Aim of the study}
The aim of the study is: 

To study temperature variations over four climatic zones
in Ghana using the Prediction of Worldwide Energy
Resources (POWER) Data.

\subsection{The Objectives of the Study}
The specific objectives of the study include the following;
\begin{itemize}
    \item To use Statistical Analysis (LR, F-Test etc.) of POWER data to study the 
seasonal variations of temperature changes over the 4 climatic zones in 
Ghana.
    \item To use Landsat data to observe the temperature profile of the hottest years on record of the  study area within the study period.
    \item To compare the POWER data with Landsat-derived LST profile maps. 
    \item To predict using the POWER data the temperature variations over the 4 Climatic zones. 
\end{itemize}


\section{Literature Review}
\textcite{bessah2022climatic} conducted periodic climate zoning in Ghana based on rainfall, relative humidity, maximum and minimum temperatures. The study adopted cluster and PCA analysis methods, leading to the classification of Ghana into Savannah, Forest, and Coastal zones based on assessed parameters. Here, a transition zone between the Savannah and the forest zone has been added.However, following the assessment of solar radiation resources from the NASA‑POWER reanalysis products \parencite{chandler2013nasa} for tropical climates in Ghana, \textcite{quansah2022assessment} showed that it is possible to expand the clean energy industry in Ghana by comparing 22 on-site measurements of sunlight radiation with that of POWER for long term reference. 

Basically, they created a detailed database of solar radiation and related parameters, focusing on short time intervals. This involves gaining a deep understanding of how solar radiation is distributed and correlated in different regions and time periods. Additionally, accurate forecasts were developed to determine the performance of solar power systems and hence assessed the technical impact of variations in solar radiation on national electric grids.

By utilizing POWER for 'Solar City' identification in Ghana, recent indications via data from the  \gls{GMET}, have provided some evidence of a general quite significant soaring of temperatures in Ghana leading to Local Warming, from about 1945 up to 1990 and hence future investigations called for \parencite{stephens1995some}.

In accordance with the second World Climate Conference held in Geneva, 1990, which advocated sustained research on global climate system changes and short-term regional analyses, \textcite{klutse2020projected} focused on projected temperature increases in Northern Ghana from 1980 to 2014.
Using Regional Climate Models (RCMs) and General Circulation Models (GCMs) to downscale climate projections from GMET data, they found significant variations in minimum air temperature: $0.5^{o}C$ under RCP 2.6 by 2080 and $2.5 ^{o}C$ under RCP 8.5 for all stations.

Along with global assessments, the study suggests that global warming is on the rise \parencite{de2012climate, mesti2015}. Statistical analysis has, however, been carried out to assess different studies. Some of which include accurate models to estimate global solar radiation which includes mean bias error \parencite{jordan1971bias}, root mean square error, mean absolute percentage error, the coefficient of determination and uncertainty at 95\%. Also, linear and non-linear regression models such as; Angström–Prescott model and Duffy and Beckman \parencite{duffie1994solar} while \parencite{kaplanis2006new} used a pocket calculator model (Jains model and a proposed S. Kaplanis model) to predict hourly global solar rays that were latter compared with measured values of this parameter over two biggest cities in Greece. 

Other known statistical methods proposed by \parencite{collares1979average} showed that these statistics can be used to calculate the heating and cooling loads of buildings. A service life model was, however, developed by \parencite{talukdar2013carbonation} to evaluate the effects of climate change on these concrete infrastructures. After the development of models, four 
categories of statistical indicators could be used to assess the performance of the developed model \parencite{gueymard2014review}. This includes Class A- The indicators of the error of individual points, the ideal value is zero for a perfect model. Class B-The indicators of overall performance, the ideal value is one for a perfect model. Class C-The indicators of distribution similitude. Class D -The qualitative indicators \parencite{lindvall1999models}.

Results of statistical modeling and predictions would later be of great importance to the World Bank's G5 Sahel report showing that the number of days with temperature above 35 degrees Celsius, considered the wet-bulb global temperature upper survivable limit for human beings, will soar to more than 40 per year by 2050. A consequence of these temperatures will be heat-induced stress and mortality, and the heat will affect the vulnerable, outdoor laborers, and poverty-affected populations the most.

Though statistics are not a Wunderwaffe \parencite{von1999misuses} to extract a wealth of information from a limited sample of observations they serve as an indispensable tool in the evaluation of limited empirical evidence. According to \parencite{katz2010statistics} upcoming challenges involved in developing statistical models to accurately capture 
and understand the characteristics of complex climate extreme events, such as heatwave temperature variability would need to consider the spatial dependence for numerical models for preferred output. This is suggested because of the sparsity of data characterized by in situ meteorological measurement devices and or setups.

\section{Methodology}
\subsection{Ghana the Study Area: Accra, Kumasi, Kete-Krachi and Wa}
\begin{figure*}[p]
  \centering
  \includegraphics[width=\textwidth]{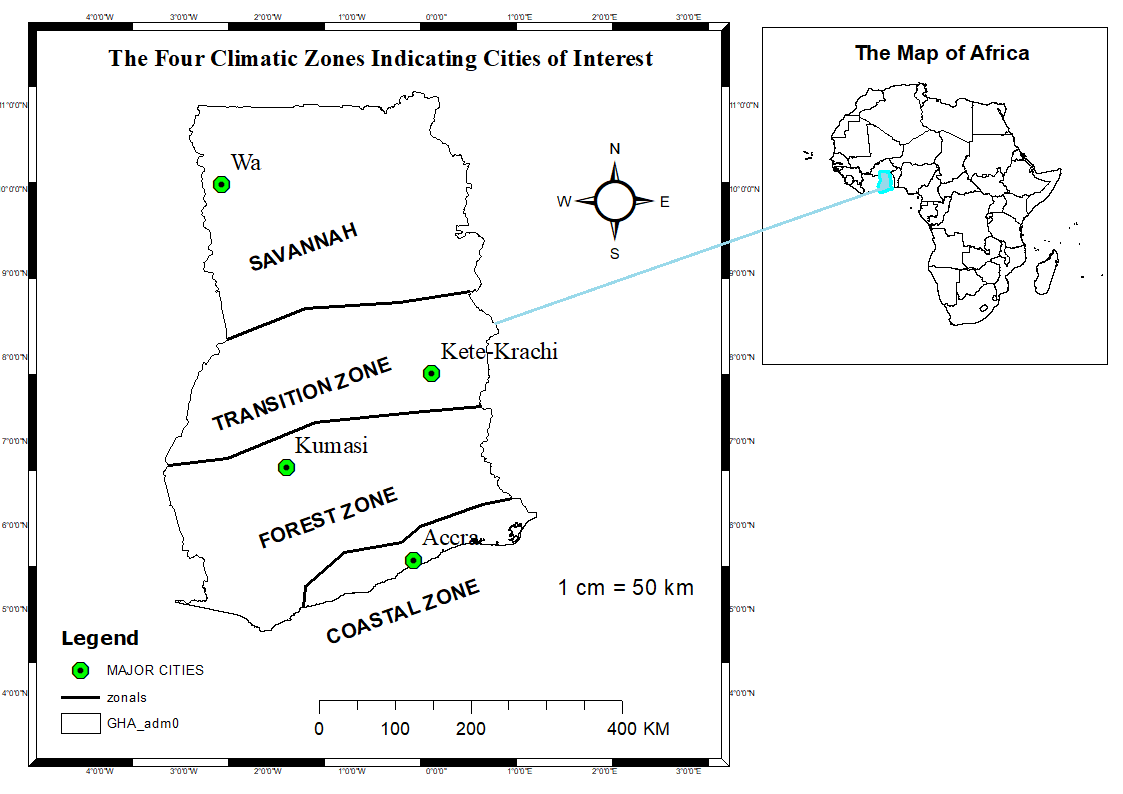} 
  \caption{The map of Ghana (in West Africa) showing the cities of interest: Accra, Kumasi, Kete-Krachi and Wa each located in their respective climatic zones: Coastal, Forest, Transition and Savannah}
  \label{fig:SA}
\end{figure*}

The study area includes four major cities in Ghana. The country is located in West Africa between latitudes $4^\circ 71'$ N and $11^\circ 20'$ N, and longitudes $1^\circ 20'$ E and $3^\circ 28'$ W. It covers a land area of about $235000, \text{km}^2$ and has a population of 30.83 million people according to the revised results of the 2021 population census. Each city is situated within a specific climatic zone as shown in Figure~\ref{fig:SA}. The Coastal Zone, characterized by cities such as Accra, experiences a maritime climate influenced by the Atlantic Ocean. In contrast, the Forest Zone represented by Kumasi is known for its equatorial rainforest climate. Kete-Krachi located in the Transition Zone is characterized by a mix of both forested and savannah landscapes, creating a transitional ecological zone while Wa, located in the Savannah Zone, showcases a semi-arid climate.

\subsection{Data Source, Data Preparation and Processing}
The original daily data for each year from 1982 to 2022, covering a total of 14975 days and 41 years for all four cities were collected from POWER's open source database provided by NASA's \gls{LaRC} POWER project funded through the NASA Earth Science/Applied Science Program as shown in Figure~\ref{fig:data_source} below. 
\begin{figure*}[p]
  \centering
  \includegraphics[width=0.7\textwidth]{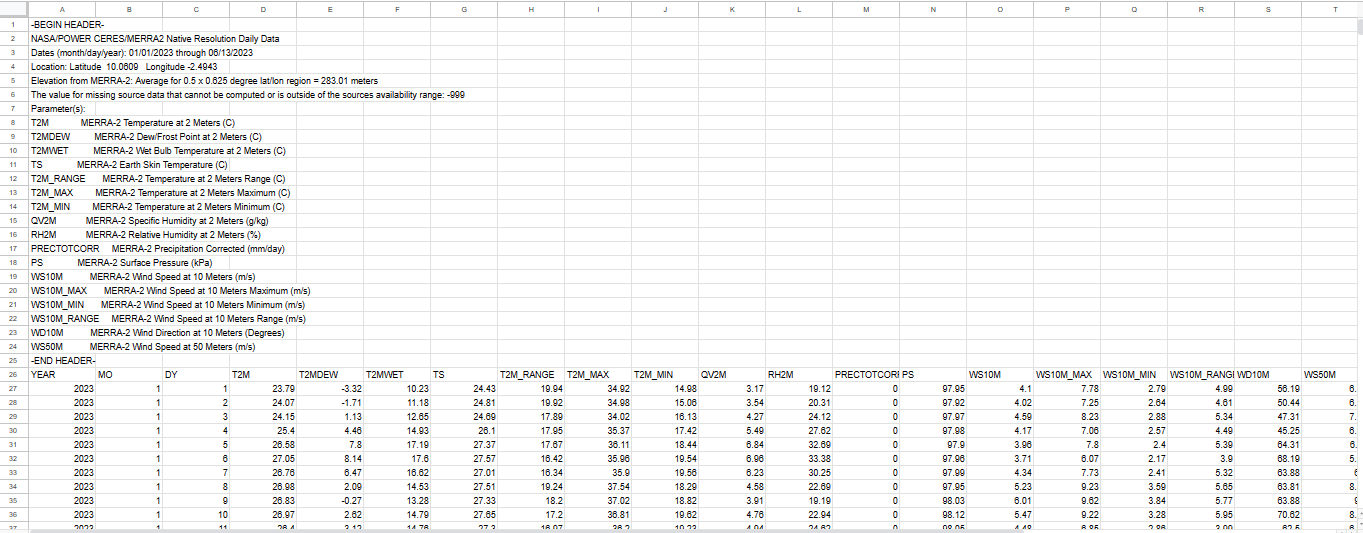}
  \caption{All Climate Parameters from NASA's Prediction of Worldwide Energy Resource Project}
  \label{fig:data_source}
\end{figure*}

The database has the advantage of being generally contiguous in time, and of being provided on a global grid having a resolution of $0.5^\circ$ latitude by $0.5^\circ$ longitude. The parameters considered for the analysis are divided into two. One for our LRT analysis and then the other sets for the XGBoost training and testing; implementing supervised ML through feature engineering. For the first part of the work, the air temperature ($^\circ$C) at the height of 2$m$ from the ground (T2M) was considered. This data has been variously abridged into yearly mean temperature since the mean of the sample serves the best predictions to be made by LRT. The following variables: \gls{T2MDEW} ($^\circ$C),  \gls{T2MWET} ($^\circ$C), \gls{TS} ($^\circ$C), \gls{T2M-Range} ($^\circ$C), \gls{QV2M} ($g/kg$), \gls{RH2M} ($\%$), \gls{PRECTOTCORR} ($mm/day$), \gls{PS} ($k Pa$), 
\gls{WS10M} ($m/s$), \gls{WS10M-Range} ($m/s$), \gls{WD10M} ($^\circ$), and \gls{WS50M} ($m/s$) all were considered in the second part of this work. 

In both cases statistical tests to evaluate our models as well as map visualizations were conducted via the following software: Jupyter Notebook is the first to mention here. It served as an open-source web application that enabled the creation and sharing of documents containing live code, visualizations, and narrative text \parencite{randles2017using, kluyver2016jupyter}. It provided an interactive computing environment where codes were executed in cells, allowing for an iterative and exploratory approach to POWER analysis. Furthermore, Jupyter Notebook served as a versatile platform for data manipulation, statistical analysis, and model development. The integration of code cells with explanatory text and visualizations in a single document facilitated the reproducibility and documentation of the research process which were later shared via Git. 

ArcMap by \textcite{chockalingam2015agriculture}, served as a powerful geographic information system (GIS) software developed by \gls{ESRI}. In this study, It provided a comprehensive set of tools and functionalities for analyzing, visualizing, and managing our  geospatial data  with its 10.7.1 version. It allowed for the manipulation and visualization of various data layers of the satellite imageries, and shapefiles, enabling the creation of informative and visually appealing maps. Though ArcMap 10.7.1's extensive toolset facilitated tasks such as cartographic design of the LST maps, these were later discarded for reasonable purposes  to the use of an alternative cloud-based LST platform (which uses \gls{GEE} plug-in) that would be addressed later, enhancing the research's geospatial data exploration and interpretation capabilities.

\begin{figure}[htbp]
  \centering
  \includegraphics[width=\textwidth]{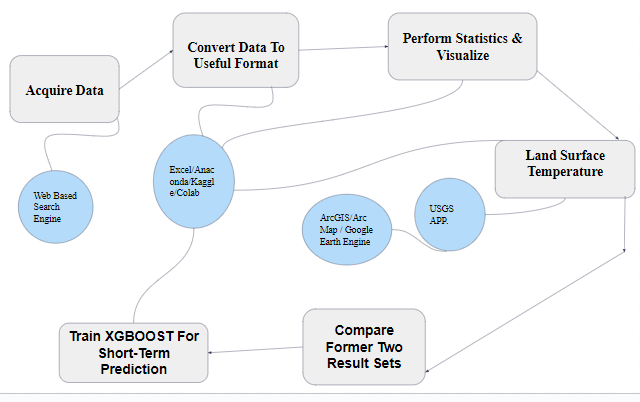}
  \caption{All Climate Parameters from NASA's Prediction of Worldwide Energy Resource Project}
  \label{fig:flow}
\end{figure}

Generally, the workflow for this project followed the following steps and shown in Figure~\ref{fig:flow}

\begin{itemize}
    \item Data was acquired from the NASA's POWER database
    \item The data was pre-processed to check for empty spaces.
    \item The data was then converted into a csv file format that can be recognized by the jupyter notebook application software.
    \item The statistical analysis which includes ANOVA test and T-tests hypothesis were formulated based on the LRT analysis. 
    \item Landsat data was then deployed from USGS for LST visualization on the RSlab global LST platform.
    \item The LRT analysis of POWER was then compared with the LST maps.
    \item The Training of the Machine Learning model then began with future predictions into mid-2023 only.
\end{itemize}

\section{Results and Discussions}
\subsection{The Statistical Analysis (LR, F-Test etc.) of POWER Data To Study The Seasonal
Variations Of Temperature Changes Over The 4 Climatic Zones In Ghana}

\begin{figure}[htbp]
    \centering
    
    \begin{subfigure}[b]{0.49\textwidth}
        \centering
        \includegraphics[width=\textwidth]{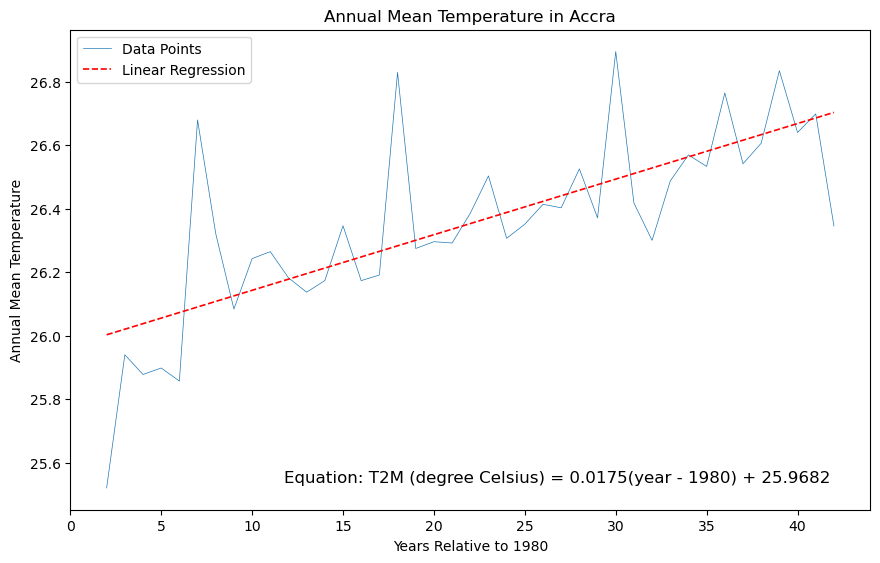}
        \caption{LRT of T2M over Accra}
        \label{fig:4.1a}
    \end{subfigure}
    \hfill
    \begin{subfigure}[b]{0.49\textwidth}
        \centering
        \includegraphics[width=\textwidth]{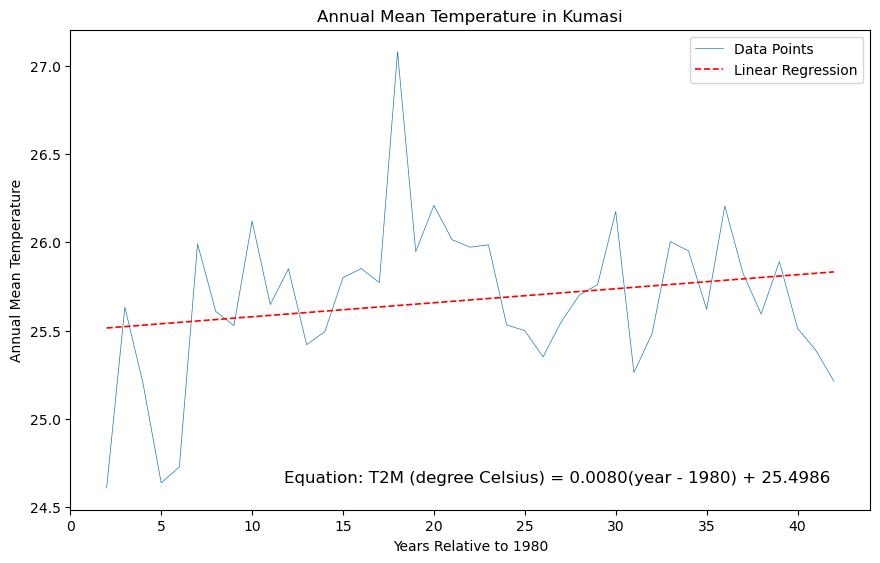}
        \caption{LRT of T2M over Kumasi}
        \label{fig:4.1b}
    \end{subfigure}
    
    \vspace{0.5cm} 
    
    \begin{subfigure}[b]{0.49\textwidth}
        \centering
        \includegraphics[width=\textwidth]{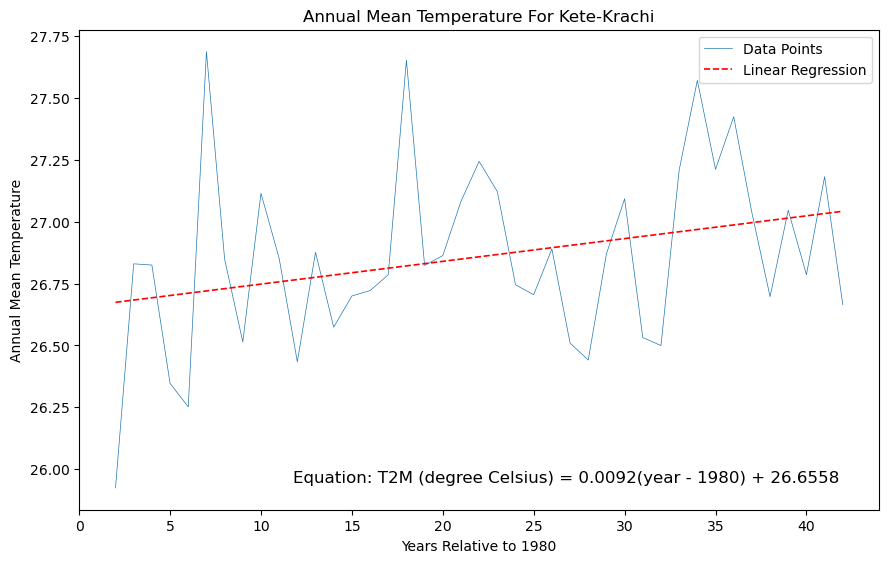}
        \caption{LRT of T2M over Kete-Krachi}
        \label{fig:4.1c}
    \end{subfigure}
    \hfill
    \begin{subfigure}[b]{0.49\textwidth}
        \centering
        \includegraphics[width=\textwidth]{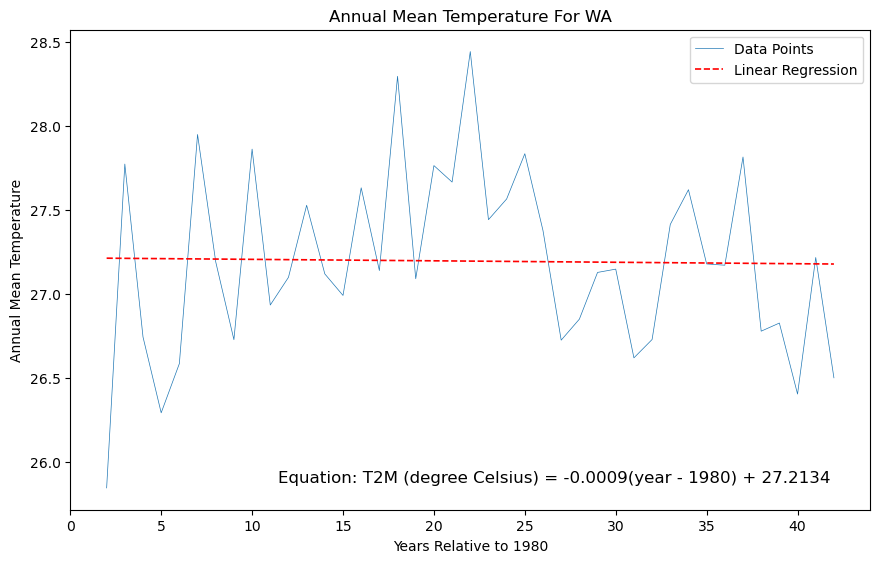}
        \caption{LRT of T2M over Wa}
        \label{fig:4.1d}
    \end{subfigure}
    
   \caption{Temporal Evolution of Annual-Mean Air Temperature 2-Meter Height: Historical Data and LRT Analysis. The x-axis spans the years 1982 to 2022 relative to 1980, capturing the long-term trends and variations.}
    \label{fig:overall}
\end{figure}

To begin with, the significance value of $\alpha$ = 0.05 was relevant for all analyses done. The first section depicts the models of LRT (represented with the dashed red lines) for each City via a scatter plot that was joined with a straight solid line instead of using isolated maker points. Thus the annual-mean air temperature at 2-meter height (T2M) over the years relative to 1980 was illustrated as shown in Figure~\ref{fig:overall}. The x-axis represents the years, while the y-axis represents the arithmetic mean of the T2M values averaged over all days within each year for all four cities. 
The LRT model for the mean temperature observed for Accra in Figure~\ref{fig:4.1a} produced, 
\begin{equation*}
    \overline{\text{T2M}} (^{\circ}\text{C}) = 0.0175 \cdot (\text{year} - 1980) + 25.9682. 
\end{equation*} As a measure of the usefulness of this LRT model, the corresponding R-squared value explained  53.14\% of the variance in the yearly mean temperatures. This is above midway between the best value of 100 and the worst value of 0. Thus, the R-squared value supports the LRT model as a useful one and that the years have an  explanatory power in predicting the yearly mean temperature for Accra's T2M. This linear dependence of the year and the yearly mean temperature provides evidence of local warming in Accra at a rate of approximately 0.018 $^{\circ}\text{C}$/year starting from the estimated 25.9682 $^{\circ}\text{C}$. The predicted LRT model varied greatly for the other three cities. Kumasi's LRT in Figure~\ref{fig:4.1b} is given as, \begin{equation*}
    \overline{\text{T2M}} 
      (^{\circ}\text{C}) = 0.0080 \cdot (\text{year} - 1980) + 25.4986. 
\end{equation*}
This model generated an R-squared value of 0.04536 indicating that approximately 4.54\% of the variance in the yearly mean temperature can be explained by the variation in the year. In other words, the year alone is not a strong predictor of the yearly mean temperature in Kumasi; other factors (atmospheric state parameters which were later looked at) or variables beyond the year contribute significantly to the variations observed. However, local warming observed in Kumasi is 0.008 $^{\circ}\text{C}$/year with an estimated starting temperature of 25.4986 $^{\circ}\text{C}$. Kete-Krachi was observed as the next warmest city after Accra having produced an LRT as shown in Figure~\ref{fig:4.1c} to be, \begin{equation*}
    \overline{\text{T2M}} (^{\circ}\text{C}) = 0.0092 \cdot (\text{year} - 1980) + 26.6558. 
\end{equation*} It is believed that local warming is happening at a rate convincingly approximately equal to 0.01 $^{\circ}\text{C}$/year starting from  26.6558 $^{\circ}\text{C}$. However, the R-squared value produced from its LRT model accounts for 91.21\% of the yearly mean temperatures being unexplained by the years in the regression model. Thus, this model has a higher R-squared value, approximately 8.78\% of the T2M variable which is 0.3\% below twice the explainable R-squared value produced for Kumasi. For the city of Wa in Figure~\ref{fig:4.1d}, the LRT for the mean temperature produced; \begin{equation*}
    \overline{\text{T2M}} (^{\circ}\text{C}) =  -0.0009\cdot (\text{year} - 1980) + 27.2134 
\end{equation*} with the corresponding Coefficient of Determination being 0.000356. This  suggests that only a very small proportion, 0.0356\% of the variance in the yearly mean temperature can be explained by the years in the model. The LRT model for the city of Wa is slightly inclining down to the right, in agreement with the much smaller negative estimated slope regression coefficient
of -0.0009 $^{\circ}\text{C}$ (of no T2M variability) / year.  starting from an estimated value of 27.2134 $^{\circ}\text{C}$ since 1980. Wa being in the Savannah Zone of Ghana is highly unexpected to exhibit such an LRT model. 

\begin{figure}[htbp]
    \centering
    
    \begin{subfigure}[b]{0.49\textwidth}
        \centering
        \includegraphics[width=\textwidth]{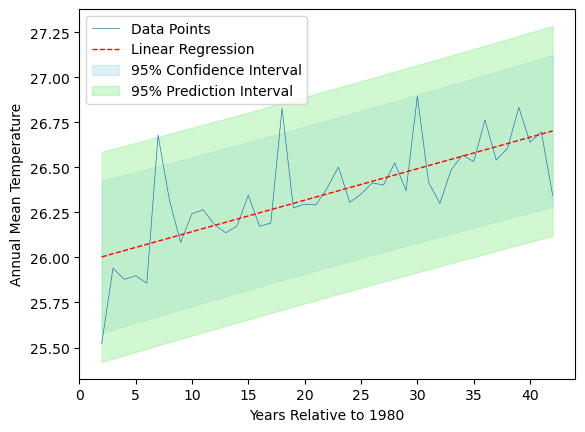}
        \caption{The Confidence \& Prediction Intervals for Accra T2M historic plot.}
    \end{subfigure}
    \hfill
    \begin{subfigure}[b]{0.49\textwidth}
        \centering
        \includegraphics[width=\textwidth]{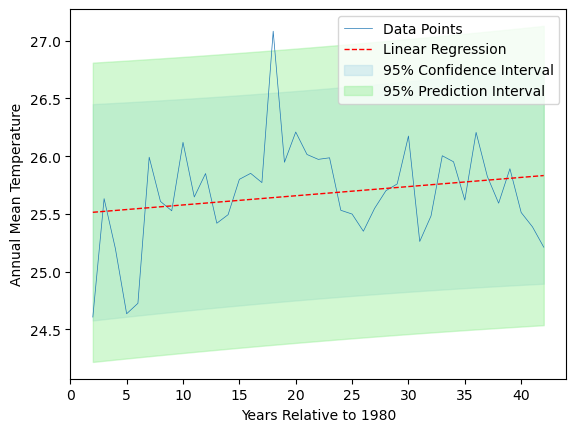}
        \caption{The Confidence \& Prediction Intervals for Kumasi T2M historic plot.}
    \end{subfigure}
    
    \vspace{0.5cm} 
    
    \begin{subfigure}[b]{0.49\textwidth}
        \centering
        \includegraphics[width=\textwidth]{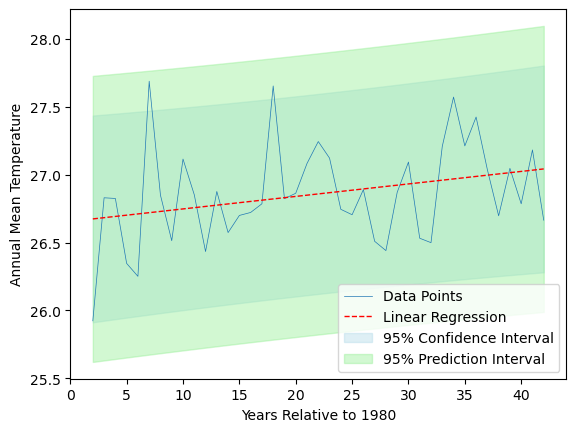}
        \caption{The Confidence \& Prediction Intervals for Kete-Krachi T2M historic plot.}
    \end{subfigure}
    \hfill
    \begin{subfigure}[b]{0.49\textwidth}
        \centering
        \includegraphics[width=\textwidth]{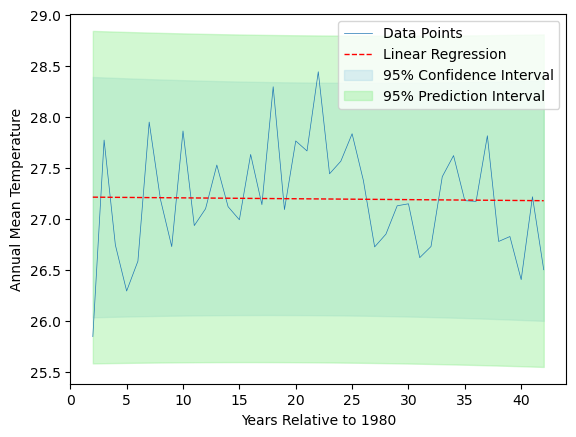}
        \caption{The Confidence \& Prediction Intervals for Wa T2M historic plot.}
    \end{subfigure}
    
    \caption{The 95\% prediction and 95\% confidence intervals of the 2-meter air temperature for all four cities. The 95\% prediction interval (PI) bounds are highlighted in pale green. The 95\% confidence interval (CI) bounds are highlighted in pale blue.}
    \label{fig:PRECONF}
\end{figure}

For the variations observed so far, the 95\% confidence interval and 95\% prediction intervals were also analyzed given in Figure~\ref{fig:PRECONF}. The 95\% confidence interval represented the range within which we were 95\% confident that the true best-fit regression line lies while the 95\% prediction interval provided the range within which we were 95\% confident that future observations will fall, given a specific value of the predictor variable (in this case the year).
A significant observation of this is to the 95\% prediction interval for Kumasi where the year 1998 was found above the 95\% prediction interval. This suggests that the observed yearly mean temperature for 1998 is significantly different from what was predicted by the regression model and scientifically proven by inspection to be the commonest warmest year on record for all four cities and correspondingly the 10th warmest year globally \parencite{noaa2020}.

\subsubsection{Hypothesis Testing}

The \gls{ANOVA} \parencite{heumann2016introduction} test to further validate our findings was carried out for all cities based on the following: (1) Null Hypothesis (H0): The LRT model has no significant effect on explaining the variation in the Mean Temperature, and 
(2) Alternative Hypothesis (HA): The LRT model has a significant effect on explaining the variation in the mean temperature. In Table \ref{tab:anova}, the ANOVA test for the mean temperature variations over Accra is shown. The term (df) is the degrees of freedom, the
term (sum\_sq) is the sum of squares, the term (mean\_sq) is the mean of squares. 
\begin{table*}[p] 
\centering
\caption{ANOVA Table for Accra T2M}
\label{tab:anova}
\begin{tabular}{cccccc}
\hline
Source & df & sum\_sq & mean\_sq & f & p\_value \\
\hline
Regression & 1.0 & 1.756237 & 1.756237 & 44.225294 & $6.498156 \times 10^{-8}$ \\
Residual & 39.0 & 1.548735 & 0.039711 & NaN & NaN \\
\hline
\end{tabular}
\end{table*}

The p\_value being $6.498156 \times 10^{-8}$ is much smaller than the arbitrarily-set target significance level of $\alpha = 0.05$. Thus, the LRT model for Accra is further supported hence we reject the null hypothesis. The reported p\_value in all cases is (1 – CFD) for the F-distribution at the obtained f-value where CFD is the cumulative
density function for the F-distribution, evaluated here with the first degree of freedom being 1 and the second degree of freedom being 39.
The ANOVA table for Kumasi is in Table~\ref{tab:anova2}.
Its associated p\_value for the f-statistics for the LRT is 0.18119. Since this p\_value is greater than the significance level of 0.05, we fail to reject the null hypothesis. 
\begin{table}[htbp]
\centering
\caption{ANOVA Table for Kumasi T2M}
\label{tab:anova2}
\begin{tabular}{cccccc}
\hline
Source & df & sum\_sq & mean\_sq & f & p\_value \\
\hline
Regression & 1.0 & 0.363165 & 0.363165 & 1.853507 & 0.18119 \\
Residual & 39.0 & 7.641423 & 0.195934  & NaN & NaN \\
\hline
\end{tabular}
\end{table}
Kete-Krachi and Wa's ANOVA Tables are displayed in Tables~\ref{tab:Ket} and ~\ref{tab:wa}
\begin{table}[htbp]
\centering
\caption{ANOVA Table for Kete-Krachi T2M}
\label{tab:Ket}
\begin{tabular}{cccccc}
\hline
Source & df & sum\_sq & mean\_sq & f & p\_value \\
\hline
Regression & 1.0 & 0.486588 & 0.486588 & 3.756572 & 0.059864 \\
Residual & 39.0 & 5.051658 & 0.129530   & NaN & NaN \\
\hline
\end{tabular}
\end{table}

\begin{table}[htbp]
\centering
\caption{ANOVA Table for Wa T2M}
\label{tab:wa}
\begin{tabular}{cccccc}
\hline
Source & df & sum\_sq & mean\_sq & f & p\_value \\
\hline
Regression & 1.0 & 0.004311 & 0.004311 & 0.013892  & 0.906778 \\
Residual & 39.0 & 12.100878 & 0.310279   & NaN & NaN \\
\hline
\end{tabular}
\end{table}
In Kete-Krachi, the p\_value of 0.059864 (which is $ \geq 0.05$
) is thought to not be of enough evidence in failing to reject the null hypothesis as in the case of Kumasi. Thus, better still there could be more reasons the null hypothesis can be rejected in favor of the alternative hypothesis - one of which could be conducting other statistical tests or considering demographic factors which result in anthropogenic activities hence high-temperature changes in this city. On the other hand, the p\_value for Wa is $ > $ than the chosen significant value therefore, we fail to reject the null hypothesis as well. 

\subsubsection{T-Test}
Understanding the significance of the b0 component and b1 component of all LRT models done so far was further carried out through the student test under the following hypothesis.
(1) For b0, Null hypothesis (H0): The mean temperature in a specific year is equal to the mean temperature in a reference year (1980). Alternative hypothesis (HA), The mean temperature in a specific year is not equal to the mean temperature in a reference year.
(2) For b1:Null hypothesis (H0): There is no linear relationship between the relative year and the mean temperature. Alternative hypothesis (HA): There is a linear relationship between the relative year and the mean temperature. The t-statistics, the upper bound and lower bound was used as an evaluation tool in this case. Table~\ref{tab:t-test} shows the output of the T-test for all four cities. 
This results now validate the fact that, without reasonable doubt it is statistically certain that the city of Accra leads the local warming in Ghana among all four cities with t-statistical values of (395.22, 6.6502) followed by Kete-Krachi with t-statistics of (224.63, 1.9382) and then Kumasi to Wa in that order. Unlike the test of ANOVA, the high values gotten from the t-statistics for Kete-Krachi do validate now that Kete-Krachi is one of the rising warmest cities making us to confidently reject the first null hypothesis of ANOVA in favor of the alternative hypothesis for Kete-Krachi. The only problem now lies in Wa but kumasi being the third warmest city in the country could be ideal.  

\begin{table*}[p] 
\centering
\caption{T-test for the significance of the LRT model (T2M)}
\makebox[\textwidth]{%
\begin{tabular}{cccccccc}
\hline
\textbf{City} & \textbf{Coefficients} & \textbf{Standard Error} & \textbf{T-Stat} & \textbf{P-value} & \textbf{Lower 95\%} & \textbf{Upper 95\%} \\
\hline
Accra & 25.9682 & 0.0657 & 395.2298 & 0.0000 & 25.8353 & 26.1011 \\
 & 0.0175 & 0.0026 & 6.6502 & 0.0000 & 0.0122 & 0.0228 \\
\\[-0.5em]
Kumasi & 25.4986 & 0.1459 & 174.7133 & 0.0000 & 25.2034 & 25.7938 \\
 & 0.0080 & 0.0058 & 1.3614 & 0.1812 & -0.0039 & 0.0198 \\
\\[-0.5em]
Kete-Krachi & 26.6558 & 0.1187 & 224.6319 & 0.0000 & 26.4158 & 26.8959 \\
 & 0.0092 & 0.0048 & 1.9382 & 0.0599 & -0.0004 & 0.0188 \\
\\[-0.5em]
Wa & 27.2134 & 0.1837 & 148.1732 & 0.0000 & 26.8419 & 27.5848 \\
 & -0.0009 & 0.0074 & -0.1179 & 0.9068 & -0.0157 & 0.0140 \\
\hline
\end{tabular}%
}
\label{tab:t-test}
\end{table*}

\subsubsection{Population Dynamics}
\begin{figure}
  \centering
  \includegraphics[width=0.7\textwidth]{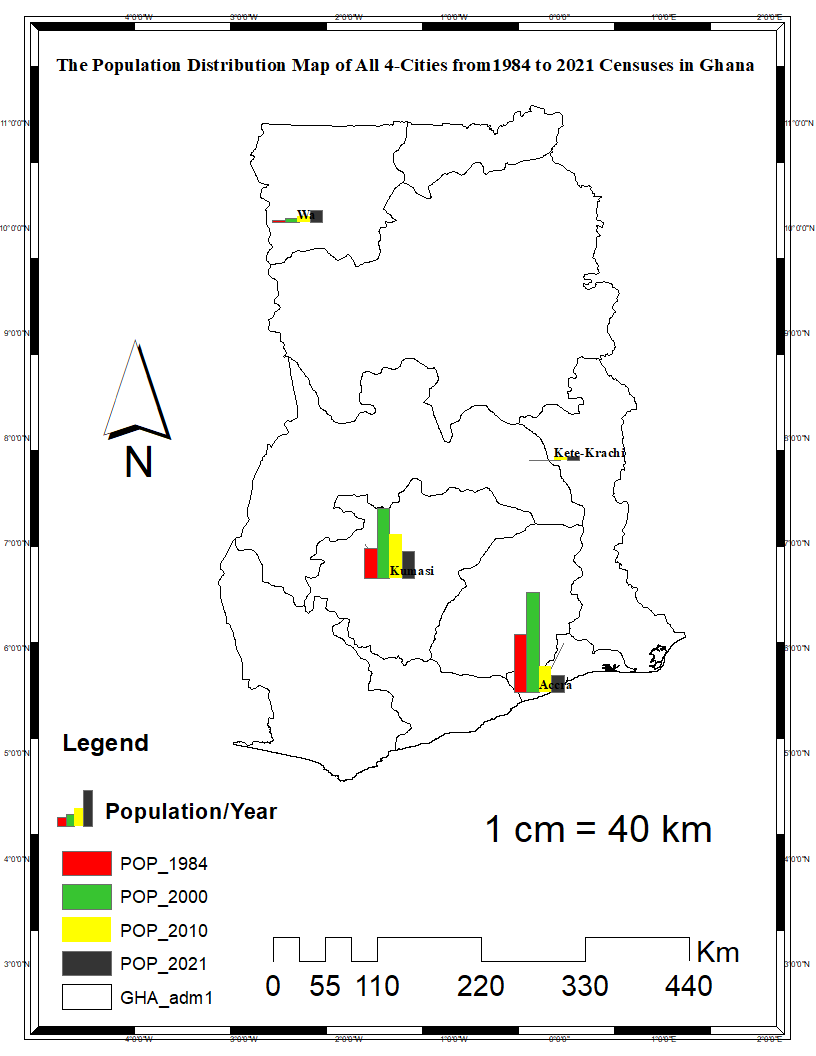} 
  \caption{Population Dynamics: 1984-2021 over Accra, Kumasi, Kete-Krachi and Wa}
  \label{fig:PD}
\end{figure}

Particularly for Wa, we have led to the proposition that the T2M variations observed may be influenced by population dynamics and its associated anthropogenic activities. To investigate this claim, an analysis was conducted on population census data spanning from 1984 to 2021 which falls within our study period for all four cities with the observed distribution map and table. These results shown in Figure~\ref{fig:PD} and  Table \ref{tab:population-dynamics}, provide a concise interpretation of the findings. 

\begin{table}[htbp]
    \centering
    \caption{Population dynamics from 1984 to 2021}
    \label{tab:population-dynamics}
    \begin{tabular}{|c|c|c|c|}
    \hline
    \textbf{City} & \textbf{1984-2000} & \textbf{2000-2010} & \textbf{2010-2021} \\
    \hline
    Accra & $+71\%\uparrow$ & $+24.8\%\uparrow$ & $-86.3\%\downarrow$ \\
    \hline
    Kumasi & $+135\%\uparrow$ & $+73.9\%\uparrow$ & $-78.2\%\downarrow$ \\
    \hline
    Kete-Krachi & N/A & N/A & $+23.7\%\uparrow$ \\
    \hline
    Wa & $+84.7\%\uparrow$ & $+6.6\%\uparrow$ & $+182\%\uparrow$ \\
    \hline
    \end{tabular}
\end{table}
\begin{itemize}
    \item Despite the recent decrease in the population of Accra, temperatures are still of great variability. This then could be the result of the industrialization of this region which results in many releases of GHGs.
    \item Kumasi being in the forest zone and having a great decrease in her population by -78\% has a well-explainable LRT model that perfectly ascertains our claims. Therefore, a decrease in local warming over Kumasi is ideal.
    \item Kete-Krachi experiencing local warming could be attributed to its increasing population. Despite the absence of previous census data for this region, the findings suggest that the current population of Kete-Krachi has a significant influence on her temperature above 2 meters in height hence its soaring data as compared to Accra. 
    \item Wa's past and current population can be thought of as a neutral agent in explaining or having rather no effect on the temperature changes in the city, regardless of an increase or decrease as a constant increase is seen throughout these years.
\end{itemize}

\subsection{Using Landsat Data To Observe The Temperature Profile Of The Hottest Years On Record Of The Study Area Within The Study Period.}
The accurate calculation and visualization of the LSTs using Landsat's Thermal Infrared Band in ArcGIS Desktop have posed challenges; due to the wide range of temperature values observed (ranging from tens to thousands of degrees Celsius), and the unavailability of data. To address this limitation and improve the LST profiles, an alternative approach was adopted by utilizing the Global Land Surface Temperature estimator available on the RSLab platform, which specifically caters to Landsat 5, 7, and 8 data while considering only the top 3 warmest years \parencite{parastatidis2017online}. The RSLab platform leverages a \gls{SC} algorithm based on the radiative transfer equation in Equation \ref{eq:Lst}, utilizing radiance-at-the-sensor data from a single band \parencite{jimenez2008revision, jimenez2014land}. 
\begin{equation}
    \label{eq:Lst}
    \text{B(LST)} = \frac{\text{L\_sen} - \text{L\_up} - \tau \cdot \big( 1 - \epsilon\big) \cdot \frac{\text{L\_down}}{\pi}}{\tau \cdot \epsilon}
\end{equation}
 B = The Planck Function \\ 
 L\_sen = The Radiance-at-the-Sensor \\
 L\_up = The Thermal Path Radiance \\ 
 L\_down = The Downwelling Irradiance \\
$\epsilon$  = The Surface Emissivity and \\
$\tau$ = The atmospheric Transmissivity.

\begin{figure*}[p]
  \centering
  \includegraphics[width=0.8\textwidth]{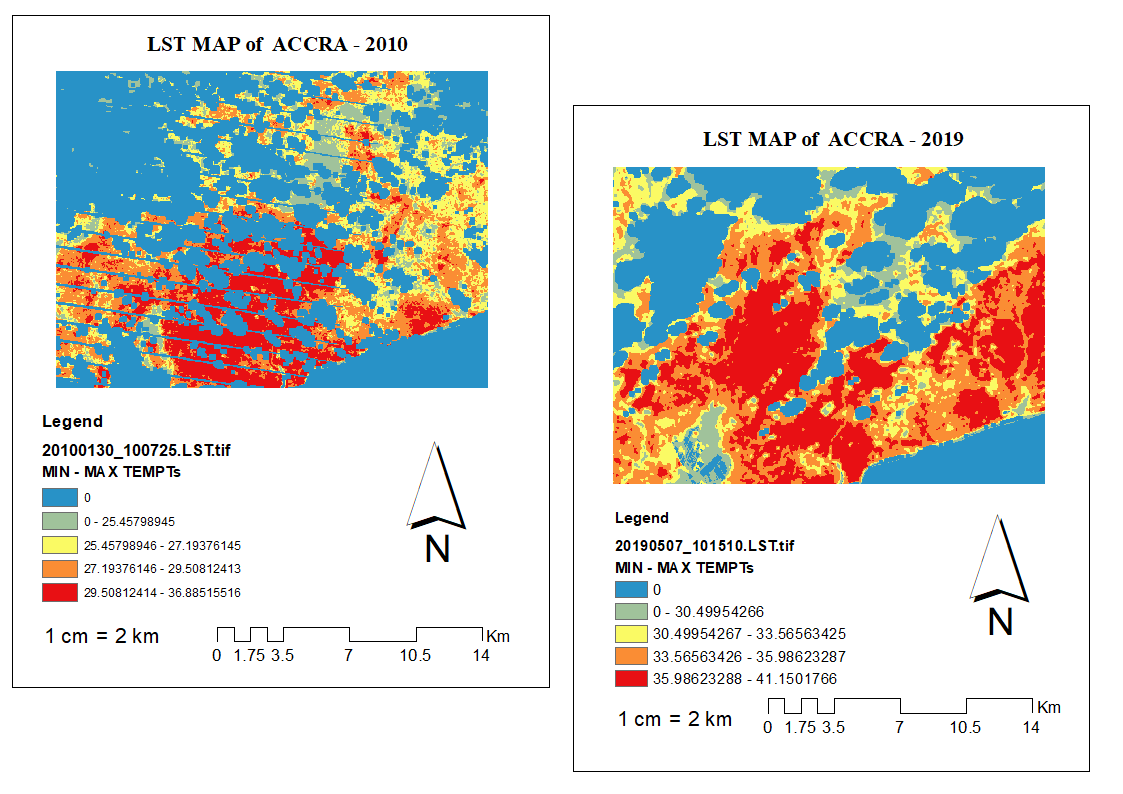} 
  \caption{LST for Accra covering 2010 to 2019}
  \label{fig:LSTACC}
\end{figure*}

Other methods of LST retrieval from Landsat data include using the radiative transfer equation in-situ radiosounding data \parencite{sobrino2004land} and the mono-window algorithm developed by \parencite{qin2001mono} using Emissivity, Transmittance and Effective Mean Atmospheric Temperature as input parameters. By integrating user preferences and employing advanced processing techniques, the platform offers enhanced LST estimation capabilities. To access LST data using the RSLab platform, users are provided with a user-friendly frontend interface that allows them to define their geographical area of interest, select the desired date range, specify the emissivity source (NDVI was used in this study), and the appropriate Landsat satellite. The user preferences are transmitted as an \gls{AJAX} request to the backend, which processes the data using the Google Earth Engine (GEE) product catalog and function. The end result was that the utilization of the RSLab platform for LST estimation has demonstrated promising results. With an accuracy of 1.52 $^\circ\text{C}$, the platform enhances the precision and reliability of LST calculations compared to previous methods from ArcMap. Below are the LST Maps of all four cities under our study for the top 3 warmest years with 1998 excluded due to unavailability of data. The LST map for Accra (Figure \ref{fig:LSTACC}) presents the temperature distribution from 2010 to 2019 (being the 1st and 2nd warmest years), providing a comprehensive view of the city's temperature profile.

\begin{figure}
  \centering
  \includegraphics[width=0.8\textwidth]{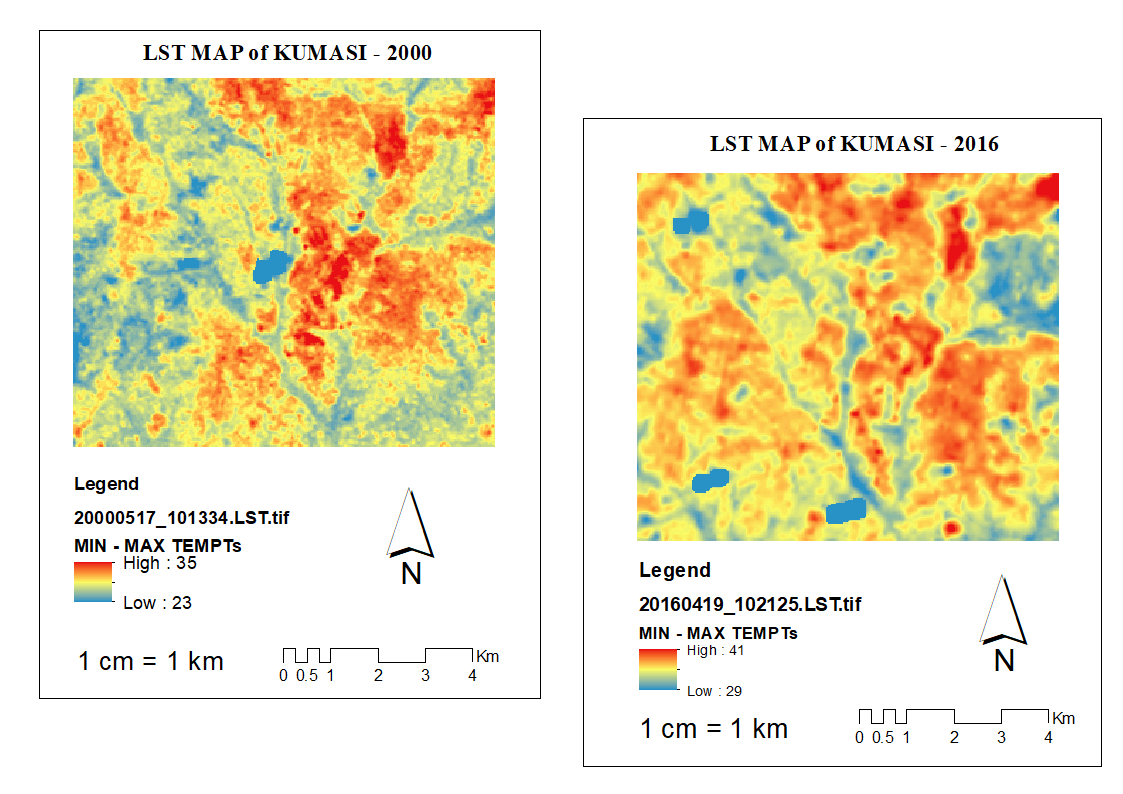} 
  \caption{LST for KUMASI covering from 2000 to 2016}
  \label{fig:LSTKUM}
\end{figure}
\begin{figure}
  \centering
  \includegraphics[width=0.8\textwidth]{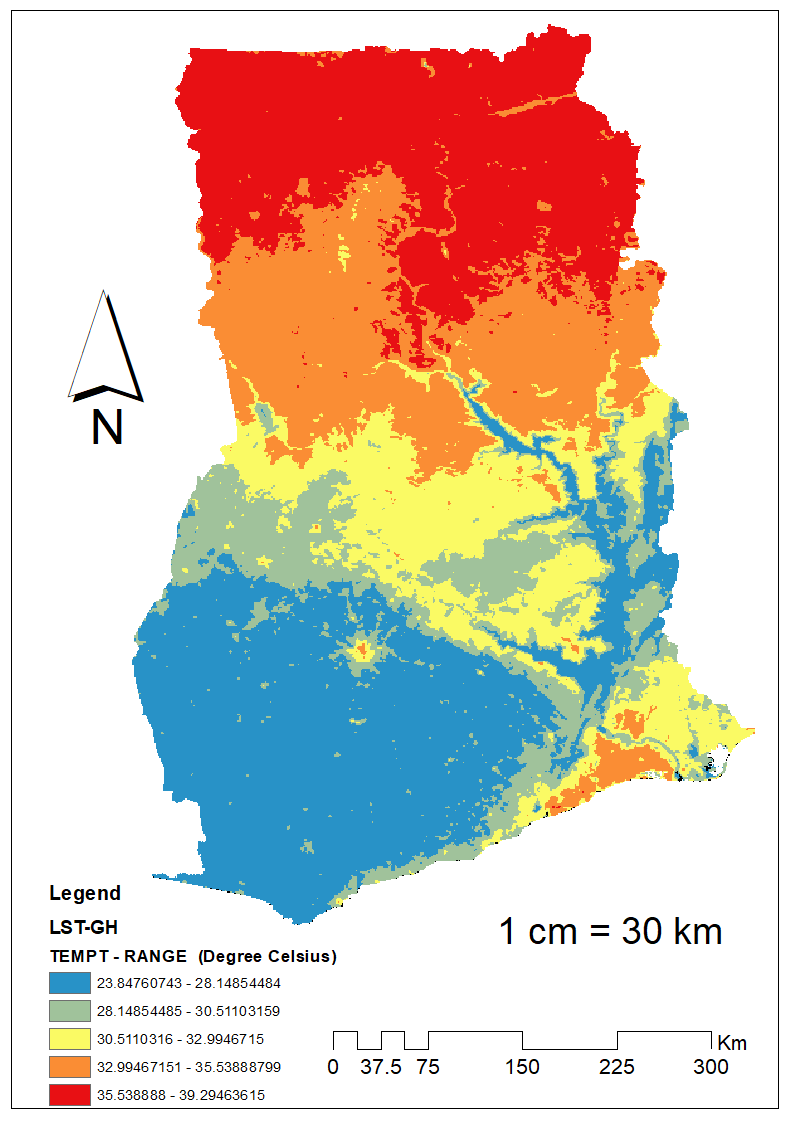} 
  \caption{Land Surface Temperature over Ghana's Climatic Zones from 1982 to 2022}
  \label{fig:LsClimatic}
\end{figure}

Similarly, the LST map for Kumasi (Figure \ref{fig:LSTKUM}) covers the years 2000 to 2016 (being the 2nd and 3rd warmest years), offering an understanding of temperature variations in that region over the hottest years.
\begin{figure}
  \centering
  \includegraphics[width=0.8\textwidth]{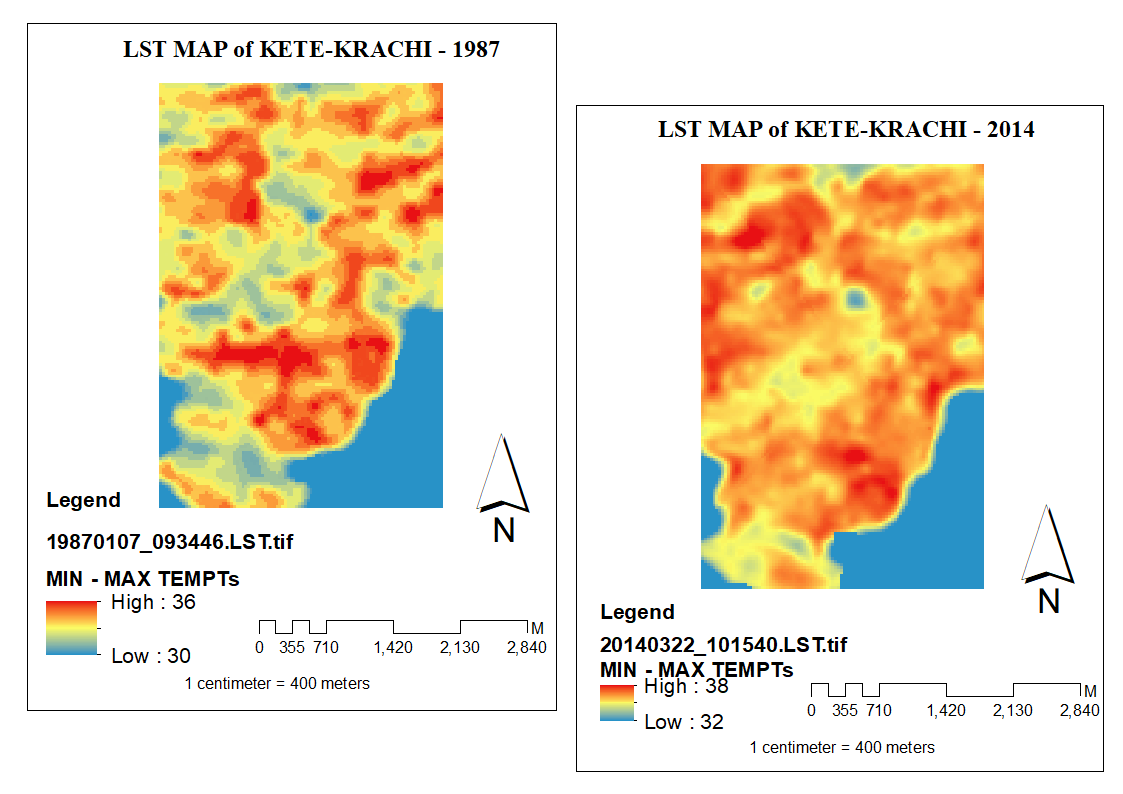} 
  \caption{LST for KUMASI covering from 1987 to 2014}
  \label{fig:LSTKET}
\end{figure}
 The LST map for Kete-Krachi (Figure \ref{fig:LSTKET}) showcases temperature patterns spanning from 1987 to 2014 (being the 1st and 3rd warmest years), aiding our analysis of long-term temperature changes.
\begin{figure*}[p]
  \centering
  \includegraphics[width=\textwidth]{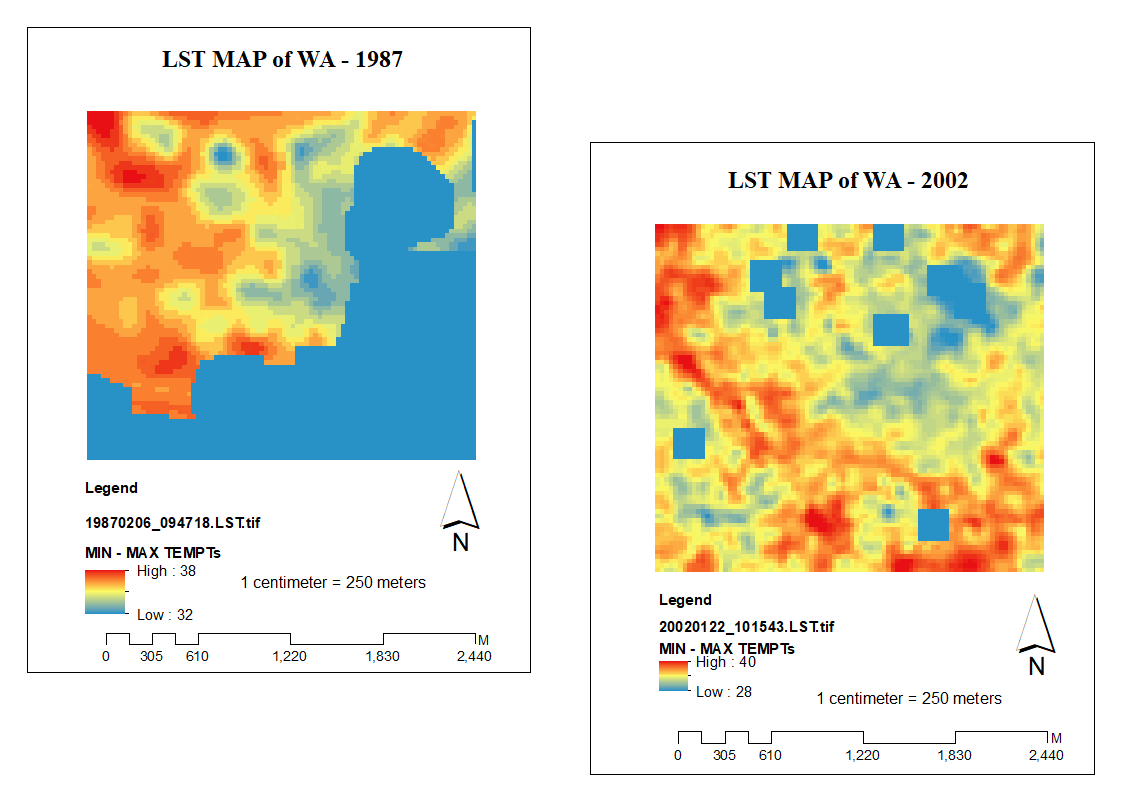} 
  \caption{LST for Wa covering from 1987 to 2002}
  \label{fig:LSTWa}
\end{figure*}
Lastly, the LST map for Wa (Figure \ref{fig:LSTWa}) provides insights into the temperature dynamics of the city during the years 1987 to 2002 (being the 3rd and 1st warmest years). Together, these LST maps can be comprehended much better while looking at the temperature variations 
 over all four climatic zones from 1982 - 2022 as shown in Figure ~\ref{fig:LsClimatic} 

In Figure ~\ref{fig:LsClimatic} it can be seen that, though Accra lies in the Coastal Zone, it has a high LST due to industrialization which is highly independent of a change in population. However, Kumasi being in the Forest zone as shown should have the observed LST / temperature changes and by the LRT model be ideal. High temperature in Kete - Krachi is also expected, unlike Wa which is experiencing lower temperatures even by a bigger projection of the results over the savannah zone.

\subsection{Comparing The POWER Data With Landsat-Derived LST Profile Maps}

One of the contributions of this work is comparing POWER with  Landsat as shown in Table~\ref{tab:compre}
\begin{table*}[p]
\centering
\caption{Comparing POWER with Landsat in Ascending order of the most temperature-varied city in Ghana.}
\makebox[\linewidth]{%
\begin{tabular}{|c|c|}
\hline
\textbf{Landsat} & \textbf{POWER} \\
\hline
Kete - Krachi & Wa \\
\cline{1-2}
Kumasi & Kumasi \\
\cline{1-2}
Wa, Accra & Kete - Krachi \\
\cline{1-2}
 & Accra \\
\cline{1-2}
\end{tabular}%
}
\label{tab:compre}
\end{table*}

While in-situ validation with POWER is yet to be conducted with GMET data of temperature variations, researches done in other parts of the world have validated POWER with ground data such as \parencite{marzouk2021assessment, monteiro2018assessment} which assessed local warming in Al Buraimi and for crop yield estimation in Brazil. Both cases have proved POWER to be reliable with in-situ measurement. However, there can be seen a great inconsistency between POWER and Landsat in telling from ascending order which city in Ghana is experiencing variations in Temperature leading to local warming. Accra maintains its position in both cases as a temperature-varied city leading to local warming. To solve this discrepancy we employed the XGBoost model to tell better predictions of the actual behavior of the POWER data.

\subsection{Prediction Using The POWER Data of The Temperature Variations Over The 4 Climatic Zones}

\begin{figure*}[p]
  \centering
  \includegraphics[width=\textwidth]{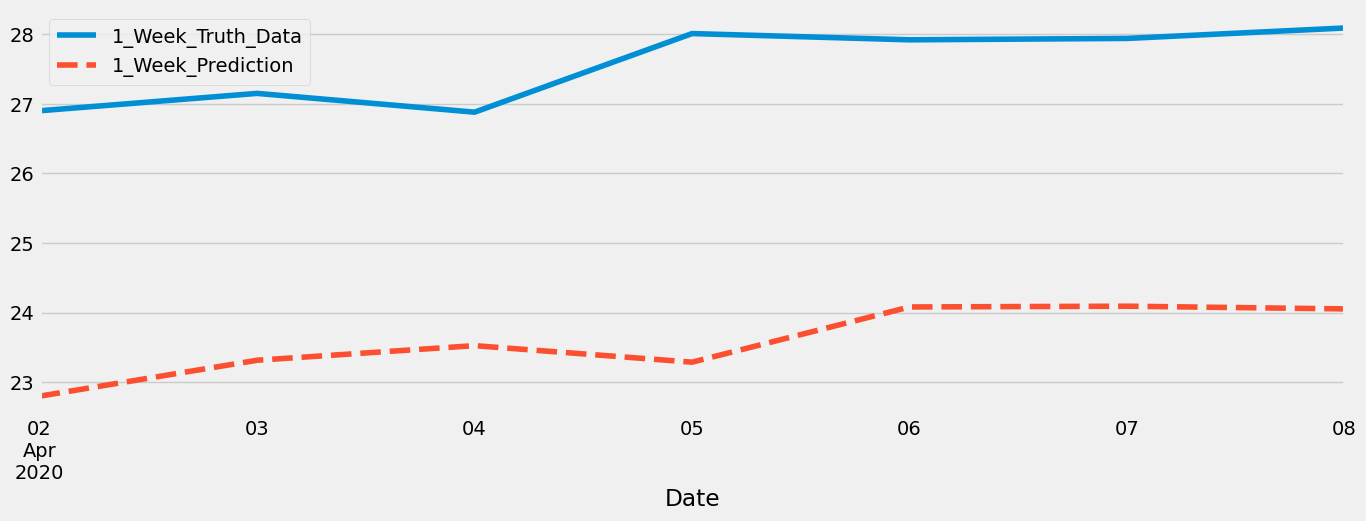} 
  \caption{A week prediction of Temperature at 2 Meters Height over Accra; from '2020-04-02' to '2020-04-08'}
  \label{fig:AccWEEK}
\end{figure*}

\begin{figure}
  \centering
  \includegraphics[width=\textwidth]{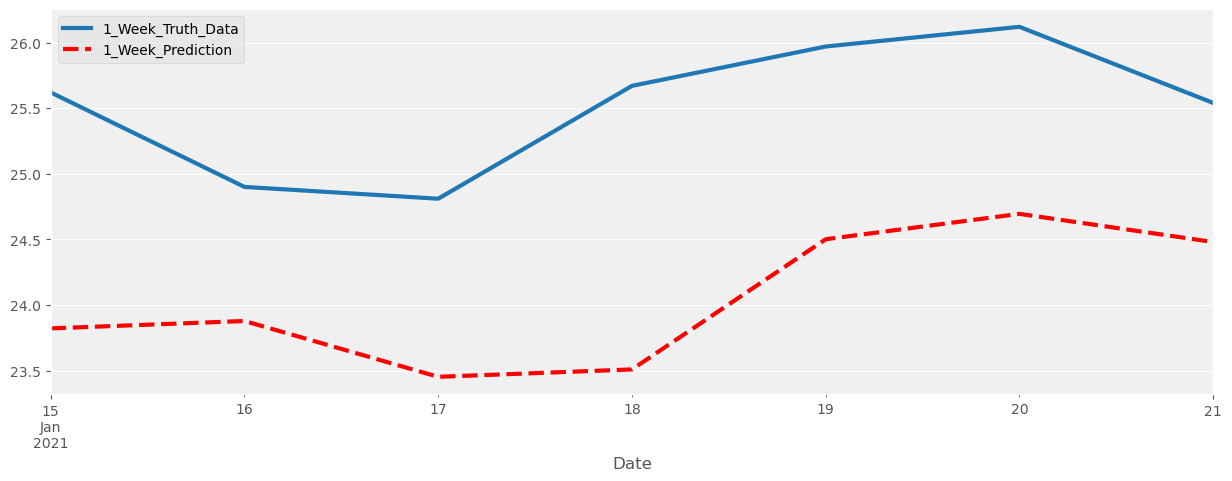} 
  \caption{A week prediction of Temperature at 2 Meters Height over Kumasi; from '2021-01-14' to '2021-01-22'.}
  \label{fig:kumpreee}
\end{figure}

XGBoost is a gradient-boosting ML model that has gained popularity in various fields, including spam classifiers, advertising systems, fraud detection, and anomalous event detection \parencite{chambers2015advanced}. It is an efficient and scalable implementation of the gradient boosting framework proposed by \parencite{friedman2001greedy, friedman2002stochastic}, which is a type of ensemble learning model. Boosting methods, such as XGBoost, combine multiple weak models to create a more accurate and robust model \parencite{kuhn2013applied}. XGBoost is similar to the traditional \gls{GBM} framework but offers better performance by using a more regularized model formalization to control overfitting \parencite{chen2022xgboost}. It has become widely used by data scientists and has been employed in more than half of the winning solutions in machine learning challenges hosted at Kaggle (He, 2016). Here, the XGBoost algorithm fits a series of small decision trees in a sequential process, where each tree is adjusted to the residuals from the previous trees. Some parameters worth knowing for training this model are: maximum depth which controls the complexity of the boosted structure and helps prevent overfitting, the learning rate determines the contribution of each tree to the model and prevents overfitting by making the boosting process more conservative, the gamma parameter specifies the minimum loss reduction required to make further partitions on leaf nodes, column and observation sampling ratios can be used to subsample variables and observations for each tree, preventing overfitting and speeding up computations, the minimum child weight parameter defines the minimum sum of instance weight required in a child, and it affects the conservatism of the algorithm, regularization, achieved through a penalty term on the weights, reduces overfitting by shrinking the consequences toward zero. Based on our model, part of the data from 1982 to 2020 was used as the training set for Accra and Kumasi. The interval 1982 to 2018 served as the training set for Wa and Kete-Krachi; we assume the model would improve as it gets to be trained on subsequent city data therefore, the time limit served as a constraint to avoid bias. The results of prediction were done Weekly, Monthly, and only yearly for best-performing City-models. Because of its less performance, the weekly predictions for Accra yielded a \gls{RMSE} of 3.582. This is illustrated by Figure~\ref{fig:AccWEEK}
This was followed by Kumasi who had an RMSE value of 1.464 as shown in Figure~\ref{fig:kumpreee}

\begin{figure}
  \centering
  \includegraphics[width=\textwidth]{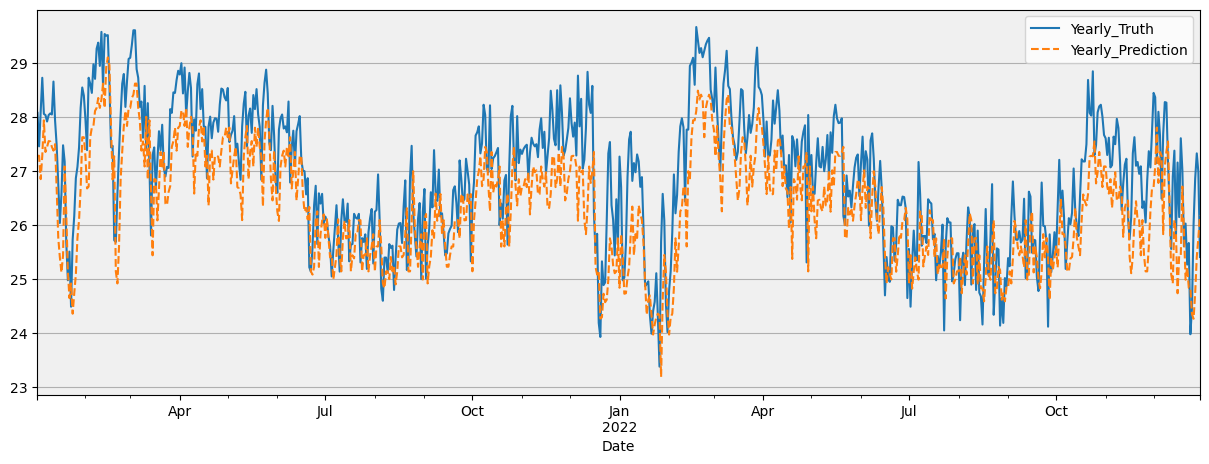} 
  \caption{Years prediction of Temperature at 2 Meters Height over Kete-Krachi; from '2021-01-01' to '2022-12-31'.}
  \label{fig:kerteee}
\end{figure}

The second-best performer was Wa with an RMSE of 1.343 with predictions of T2M being done from 2019-03-01 to 2021-04-18. The XGBoost has therefore improved so well that it outperforms itself in the previous projections made for Kumasi and Accra. We illustrate the predictions for Wa in Figure~\ref{fig:wareee} below. Though Wa has a huge unexplainable phenomenon causing constant decreasing temperatures it is confidently thought that with huge data from more atmospheric parameters, we can train and learn from these phenomena and better tell what factors contribute to the decreasing temperatures in Wa. 
\begin{figure}
  \centering
  \includegraphics[width=\textwidth]{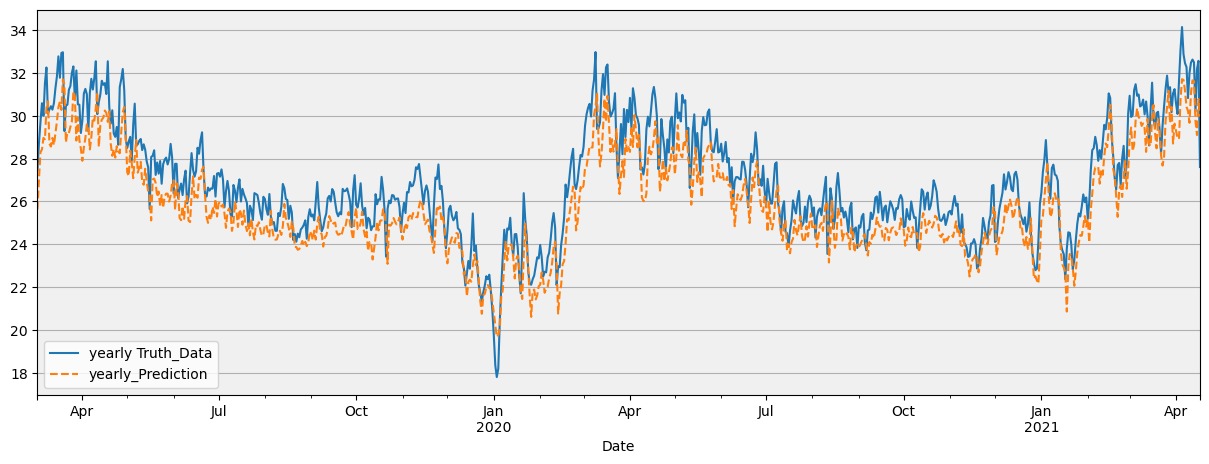} 
  \caption{Two Years prediction of Temperature at 2 Meters Height over Wa; from '2019-03-01' '2021-04-18'.}
  \label{fig:wareee}
\end{figure}
Kete Krachi produced the best T2M predictions while giving off an RMSE of 0.9876. This is demonstrated in Figure~\ref{fig:kerteee}. In addition, the best-predicted dates for Kete-Krachi are shown in Table~\ref{tab:RMSEs} with their respective RMSEs.

\begin{table}[htbp]
\centering
\caption{Root Mean Square Error (RMSE) Values}
\begin{tabular}{|c|c|}
\hline
\textbf{Date} & \textbf{RMSE Value} \\
\hline
2021-07-13 & 0.001320 \\
2019-10-27 & 0.002269 \\
2020-07-24 & 0.003424 \\
2018-05-21 & 0.003924 \\
2019-02-10 & 0.003939 \\
\hline
\end{tabular}
\label{tab:RMSEs}
\end{table}

\subsubsection{MID 2023}
\begin{figure}[htbp]
    \centering
    
    \begin{subfigure}[b]{0.49\textwidth}
        \centering
        \includegraphics[width=\textwidth]{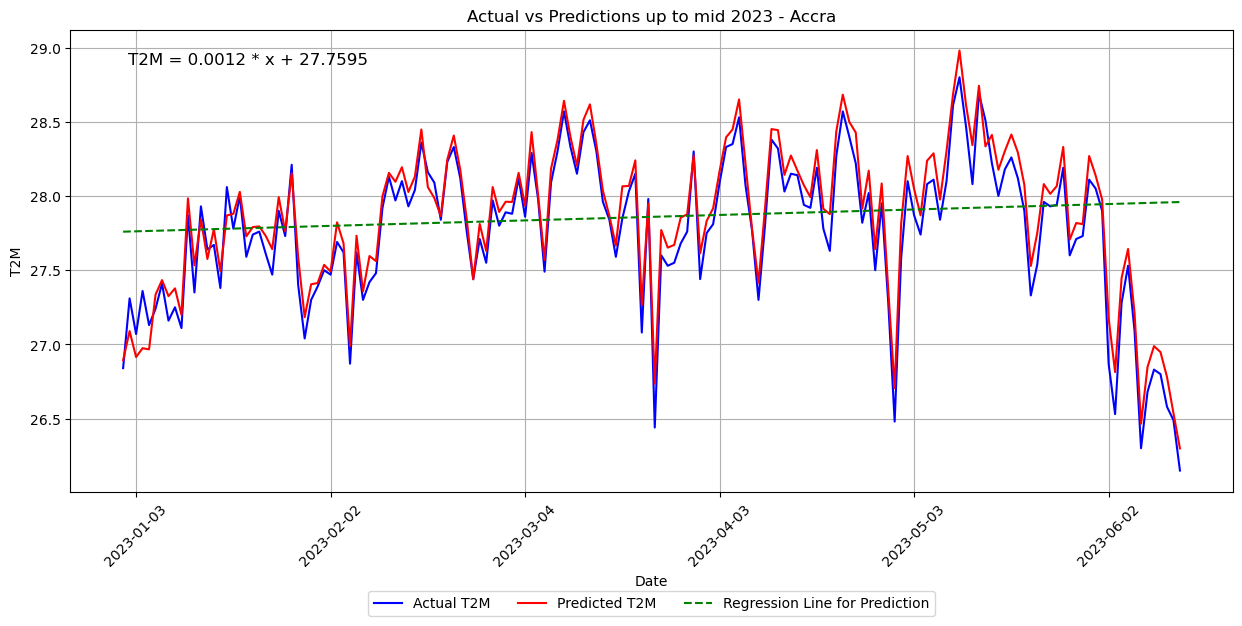}
        \caption{Mid 2023 Predictions of T2M over Accra}
        \label{fig:20231}
    \end{subfigure}
    \hfill
    \begin{subfigure}[b]{0.49\textwidth}
        \centering
        \includegraphics[width=\textwidth]{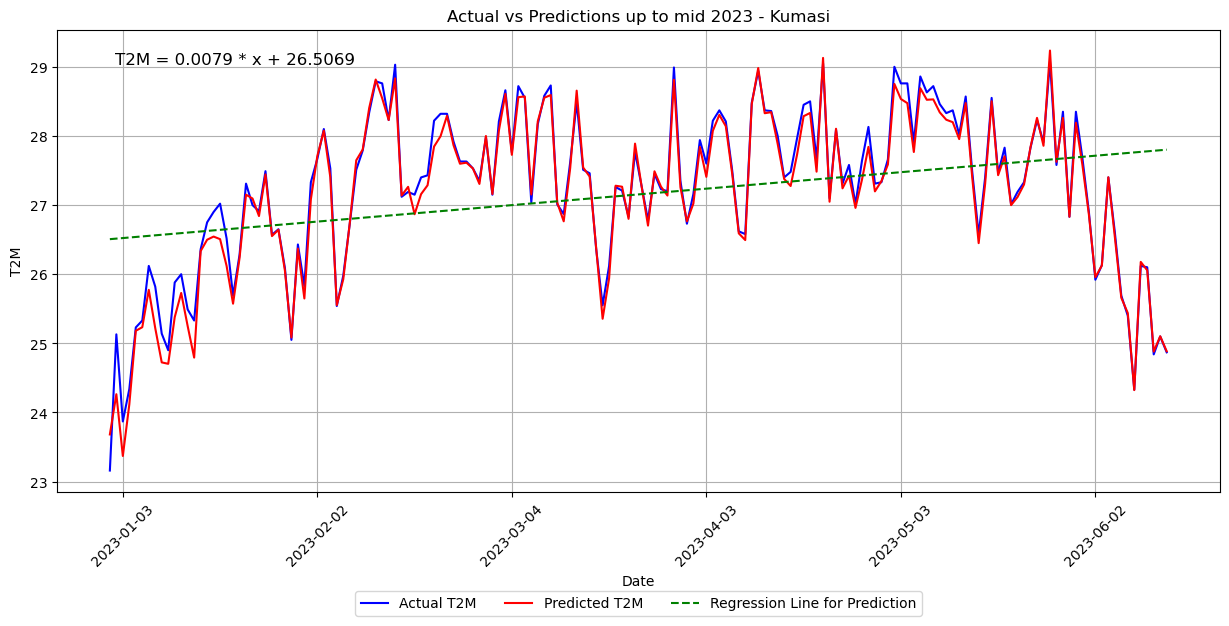}
        \caption{Mid 2023 Predictions of T2M over Kumasi}
        \label{fig:20232}
    \end{subfigure}
    
    \vspace{0.5cm} 
    
    \begin{subfigure}[b]{0.49\textwidth}
        \centering
        \includegraphics[width=\textwidth]{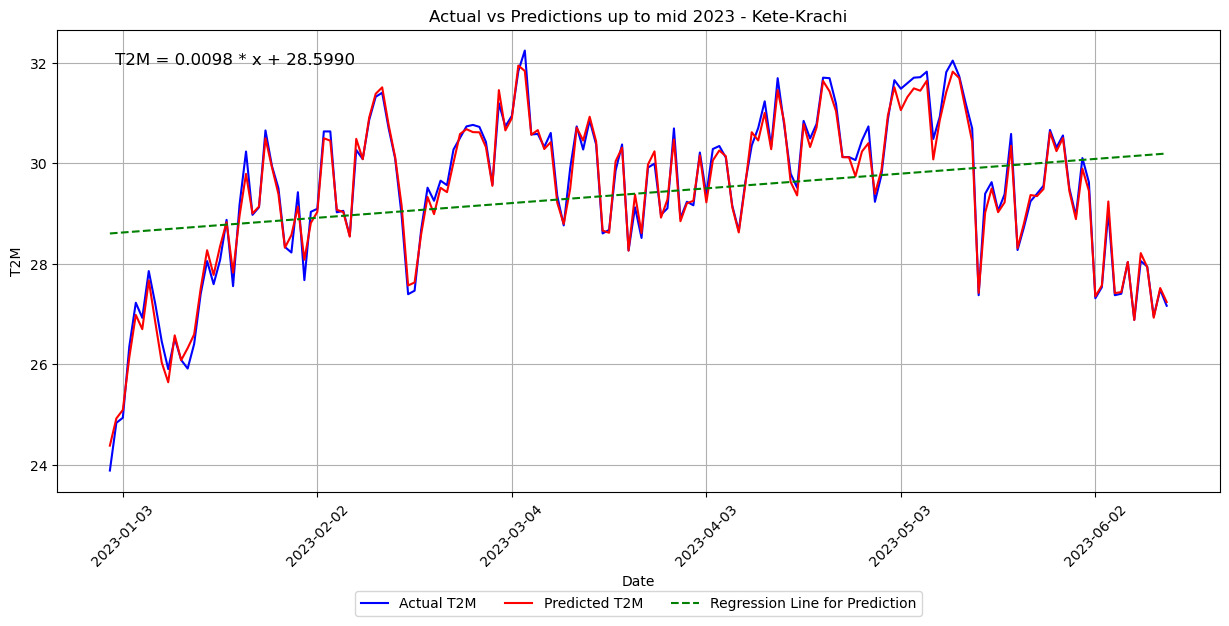}
        \caption{Mid 2023 Predictions of T2M over Kete-Krachi}
        \label{fig:20233}
    \end{subfigure}
    \hfill
    \begin{subfigure}[b]{0.49\textwidth}
        \centering
        \includegraphics[width=\textwidth]{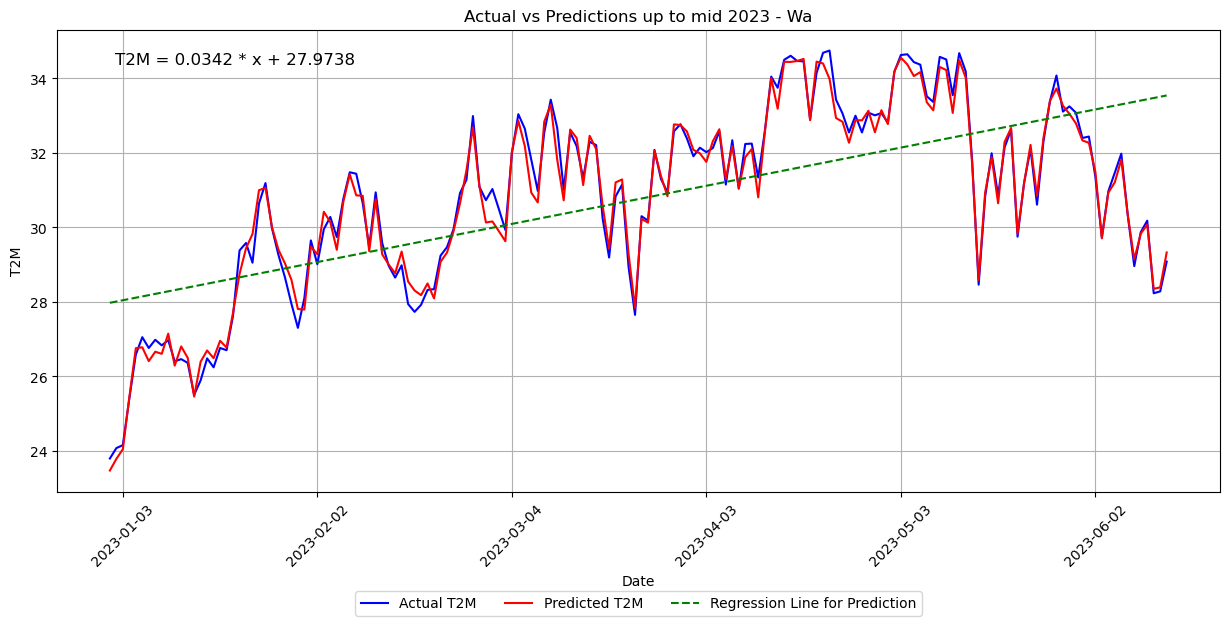}
        \caption{Mid 2023 Predictions of T2M over Wa}
        \label{fig:20234}
    \end{subfigure}
    
   \caption{Mid 2023 Predictions of T2M (Temperature) in all four Cities in Ghana}
    \label{fig:overall2}
\end{figure}

\begin{figure*}[p]
  \centering
  \includegraphics[width=\textwidth]{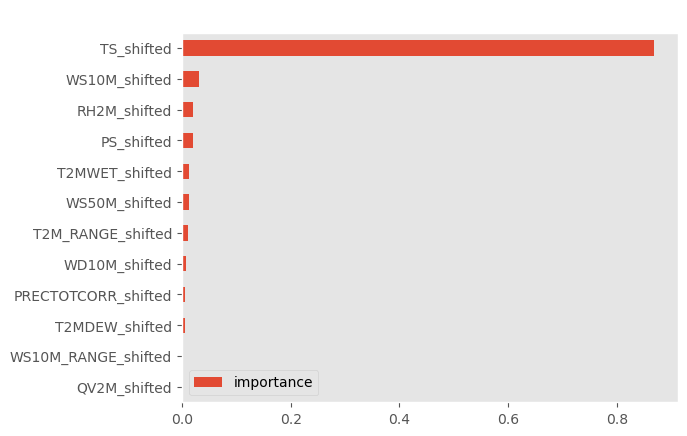} 
  \caption{The Importance level of each Climatic Variable used in Feature Engineering while training the XGBoost.}
  \label{fig:Features}
\end{figure*}

The mid-2023 prediction of mean temperature at 2 meters above the surface in these four cities is a confirmation of the projections by the XGBoost model verifying the better performance of Wa's temperature variability. This prediction of the constant high temperatures, particularly in Wa was not well explained by the LRT and though the LST showed a gradual change in this, there was always a high expectation that temperatures are significantly changing in Wa which could lead to local warming in the city.
From Figure~\ref{fig:overall2} the predictions of Wa which is Figure~\ref{fig:20234} indicates that in mid 2023, T2M is changing at a rate of 0.034$^{\circ}\text{C}$ / mid-year and the mean temperature is 30.76 $^{\circ}\text{C}$. 
Figure~\ref{fig:20233} indicates that the rate of change in T2M over Kete-Krachi in mid 2023 is 0.0098$^{\circ}\text{C}$ / mid-year and the mean temperature being 29.39 $^{\circ}\text{C}$. Accra however, is still on record high without regard to the short term analysis of the 2023 mid-year variabilities. It has a mean mid-year temperature of 27.86 $^{\circ}\text{C}$ that is changing at a rate of 0.0012$^{\circ}\text{C}$ /mid-year 
The rate of change in T2M for Kumasi is 0.0079$^{\circ}\text{C}$ / mid-year and mean mid-year temperature of 27.15 $^{\circ}\text{C}$. The Figure~\ref{fig:Features}  demonstrates which parameters were used most or less the  XGBoost  model. It can be seen that the model used the Earth's Skin Temperature about 90\% of the time to make its predictions as shown in . This confirms how temperature increase in geolocation could be influenced by the direct thermal energy coming from the land, water bodies, and other man-made features that are good absorbers and bad reflectors of thermal infrared energy.

\chapter{}
\section{Conclusion For The Study}
Using a number of statistical tools (linear regression, ANOVA, and t-test), the freely available contiguous data from NASA (under the POWER project) for air temperature at a height of 2 meters above the ground for four cities in Ghana was investigated. A total of 14,975 data points corresponding to 14,975 days covering the period from 01/January/1982 to 31/December/2022 (inclusive) produced the following: only 3 cities are experiencing Temperature Variation leading to local warming at a minimal rate for the long term study according to NASA's POWER LRT analysis. In decreasing order of change, Accra experiences temperature of 0.018 $^{\circ}\text{C}$/year, Kete-Krachi as the second locally warmed city experiences warming due to temperature variability at a rate of 0.01 $^{\circ}\text{C}$/year and Kumasi's local warming attributed to temperature changes is happening at a rate of 0.008 $^{\circ}\text{C}$/year. For a visual validation of these results, the method of geospatial analysis of LST maps derived from Landsat 5, 7, and 8 via the RSLab platform was a plus to this work and further confirming the very slow change of T2M in Wa over the years. Population dynamics over all 4 zones do not provide any insight into the temperature variabilities observed. The XGBoost Machine Learning model's short-term predictions of local warming, however, enhanced our understanding of temperature variabilities in all regions as ideal to their respective climatic zones within a short time projection. Wa tends to top the list as a high-temperature varied city in the Savannah zone followed by Kete-Krachi in the Transition Zone, then Accra (which remains heavily industrialized), and Kumasi in the Forest Zone.  

\subsection{Contribution To Knowledge And Recommendations}

The study contributed to knowledge by analyzing the Long Term study of Temperature Variations of POWER for Short Term Prediction unlike other methods that used Short Term study of in-situ data (such as GMET) for Long Term Predictions. It is believe this is possible due the Robustness of their Algorithms like GCMs and RCMs. In addition, it would be recommended based on the results derived here that, Machine Learning and artificially intelligent models should be incorporated into Large Climate Models (like GCMs) and even in-situ data collectors / devices to make inferences and Learn from the data being collected and studied. Secondly, the Skin Temperature of the Earth should be considered in making temperature predictions for local warming anywhere in the world. It is, therefore, important that, further research would focus on actually validating these results for a continental study that has been done with in-situ data by comparing it with that of POWER. 




\medskip
\cleardoublepage
\phantomsection
\addcontentsline{toc}{chapter}{Reference List}
\printbibliography[heading=Reference List]

@inproceedings{fourier1824remarques,
  title={Remarques g{\'e}n{\'e}rales sur les temp{\'e}ratures du globe terrestre et des espaces plan{\'e}taires},
  author={Fourier, Joseph},
  booktitle={Annales de Chemie et de Physique},
  volume={27},
  pages={136--167},
  year={1824}
}

@incollection{ipcc2021technical,
  title = {Technical Summary},
  author = {Arias, P.A. and et al.},
  booktitle = {Climate Change 2021: The Physical Science Basis. Contribution of Working Group I to the Sixth Assessment Report of the Intergovernmental Panel on Climate Change},
  editor = {Masson-Delmotte, V. and Zhai, P. and Pirani, A. and Connors, S.L. and Péan, C. and Berger, S. and Caud, N. and Chen, Y. and Goldfarb, L. and Gomis, M.I. and Huang, M. and Leitzell, K. and Lonnoy, E. and Matthews, J.B.R. and Maycock, T.K. and Waterfield, T. and Yelekçi, O. and Yu, R. and Zhou, B.},
  publisher = {Cambridge University Press},
  address = {Cambridge, United Kingdom and New York, NY, USA},
  year = {2021},
  pages = {33--144},
  doi = {10.1017/9781009157896.002},
}

@book{sorokhtin2007global,
  title={Global warming and global cooling: evolution of climate on Earth},
  author={Sorokhtin, Oleg Georgievich and Khilyuk, Leonid F and Chilingarian, GV},
  year={2007},
  publisher={Elsevier}
}

@article{shaftel2021overview,
  title={Overview: Weather, global warming and climate change},
  author={Shaftel, Holly and Callery, S and Jackson, R and Bailey, D},
  journal={Climate Change: Vital Signs of the Planet. https://climate. nasa. gov/resources/globalwarming-vs-climate-change/(accessed 21 June 2021)},
  year={2021}
}

@article{lindvall1999models,
  title={Models for environmental actions on concrete structures},
  author={Lindvall, A and Nilsson, LO and H{\"o}gberg, A and others},
  journal={DuraCrete document BE95-1347},
  volume={3},
  year={1999}
}

@article{dines1917heat,
  title={The heat balance of the atmosphere},
  author={Dines, WH},
  journal={Quarterly Journal of the Royal Meteorological Society},
  volume={43},
  number={182},
  pages={151--158},
  year={1917},
  publisher={Wiley Online Library}
}

@article{donohoe2013seasonal,
  title={The seasonal cycle of atmospheric heating and temperature},
  author={Donohoe, Aaron and Battisti, David S},
  journal={Journal of Climate},
  volume={26},
  number={14},
  pages={4962--4980},
  year={2013},
  publisher={American Meteorological Society}
}

@article{kolio2014durability,
  title={Durability demands related to carbonation induced corrosion for Finnish concrete buildings in changing climate},
  author={K{\"o}li{\"o}, Arto and Pakkala, Toni A and Lahdensivu, Jukka and Kiviste, Mihkel},
  journal={Engineering structures},
  volume={62},
  pages={42--52},
  year={2014},
  publisher={Elsevier}
}

@article{saha2014urban,
  title={Urban scale mapping of concrete degradation from projected climate change},
  author={Saha, Mithun and Eckelman, Matthew J},
  journal={Urban Climate},
  volume={9},
  pages={101--114},
  year={2014},
  publisher={Elsevier}
}

@article{asante2014climate,
  title={Climate change and variability in Ghana: Stocktaking},
  author={Asante, Felix A and Amuakwa-Mensah, Franklin},
  journal={Climate},
  volume={3},
  number={1},
  pages={78--101},
  year={2014},
  publisher={MDPI}
}

@misc{worldbank2014,
  author = {{World Bank}},
  title = {The Costs to Developing Countries of Adapting to Climate Change: New Methods and Estimates},
  howpublished = {Available online},
  url = {http://www.worldbank.org/eacc},
  note = {(accessed on 25th January 2014)},
  year = {2014}
}

@book{stocker2013climate,
  title = {Climate Change 2013: The Physical Science Basis. Working Group I Contribution to the Fifth Assessment Report of the Intergovernmental Panel on Climate Change - Abstract for decision-makers; Changements climatiques 2013. Les éléments scientifiques. Contribution du groupe de travail I au cinquième rapport d'évaluation du groupe d'experts intergouvernemental sur l'évolution du climat - Résumé à l'intention des décideurs},
  author = {Stocker, T. F. and others},
  year = {2013},
  publisher = {Cambridge University Press},
}

@techreport{chandler2013nasa,
  title={NASA prediction of worldwide energy resource high resolution meteorology data for sustainable building design},
  author={Chandler, William S and Hoell, James M and Westberg, David and Zhang, Taiping and Stackhouse Jr, Paul W},
  year={2013}
}

@inproceedings{chen2016xgboost,
  title={Xgboost: A scalable tree boosting system},
  author={Chen, Tianqi and Guestrin, Carlos},
  booktitle={Proceedings of the 22nd acm sigkdd international conference on knowledge discovery and data mining},
  pages={785--794},
  year={2016}
}

@article{zhang2007global,
  title={A global perspective: NASA's prediction of worldwide energy resources (POWER) project},
  author={Zhang, Taiping and Stackhouse Jr, Paul W and Chandler, William S and Hoell, James M and Westberg, David and Whitlock, Charles H},
  year={2007}
}

@article{quansah2022assessment,
  title={Assessment of solar radiation resource from the NASA-POWER reanalysis products for tropical climates in Ghana towards clean energy application},
  author={Quansah, Alfred Dawson and Dogbey, Felicia and Asilevi, Prince Junior and Boakye, Patrick and Darkwah, Lawrence and Oduro-Kwarteng, Sampson and Sokama-Neuyam, Yen Adams and Mensah, Patrick},
  journal={Scientific Reports},
  volume={12},
  number={1},
  pages={10684},
  year={2022},
  publisher={Nature Publishing Group UK London}
}

@article{stephens1995some,
  title={Some indications of global warming in Ghana},
  author={Stephens, CE},
  journal={Environmental conservation},
  volume={22},
  number={2},
  pages={174--175},
  year={1995},
  publisher={Cambridge University Press}
}

@article{klutse2020projected,
  title={Projected temperature increases over northern Ghana},
  author={Klutse, Nana Ama Browne and Owusu, Kwadwo and Boafo, Yaw Agyeman},
  journal={SN Applied Sciences},
  volume={2},
  pages={1--14},
  year={2020},
  publisher={Springer}
}

@article{de2012climate,
  title={Climate change, agriculture, and foodcrop production in Ghana},
  author={De Pinto, Alessandro and Demirag, Ulac and Haruna, Akiko},
  year={2012},
  publisher={International Food Policy Research Institute (IFPRI)}
}

@misc{mesti2015,
  title = {{Ghana's Third National Communication to the UNFCCC}},
  author = {{Ministry of Environment, Science, Technology and Innovation (MESTI)}},
  year = {2015},
  howpublished = {Retrieved from UNFCCC website},
  url = {http://unfccc.int/resource/docs/natc/ghanc3.pdf}
}

@book{jordan1971bias,
  title={Bias and mean square error in experimental designs},
  author={Jordan Filho, Leon},
  year={1971},
  publisher={Iowa State University}
}

@article{duffie1994solar,
  title={Solar engineering of thermal processes},
  author={Duffie, John A and Beckman, William A and Worek, WM},
  year={1994}
}

@article{kaplanis2006new,
  title={New methodologies to estimate the hourly global solar radiation; comparisons with existing models},
  author={Kaplanis, SN},
  journal={Renewable Energy},
  volume={31},
  number={6},
  pages={781--790},
  year={2006},
  publisher={Elsevier}
}

@article{collares1979average,
  title={The average distribution of solar radiation-correlations between diffuse and hemispherical and between daily and hourly insolation values},
  author={Collares-Pereira, Manuel and Rabl, Ari},
  journal={Solar energy},
  volume={22},
  number={2},
  pages={155--164},
  year={1979},
  publisher={Elsevier}
}

@article{talukdar2013carbonation,
  title={Carbonation in concrete infrastructure in the context of global climate change: Development of a service lifespan model},
  author={Talukdar, Sudip and Banthia, N},
  journal={Construction and Building Materials},
  volume={40},
  pages={775--782},
  year={2013},
  publisher={Elsevier}
}

@article{gueymard2014review,
  title={A review of validation methodologies and statistical performance indicators for modeled solar radiation data: Towards a better bankability of solar projects},
  author={Gueymard, Christian A},
  journal={Renewable and Sustainable Energy Reviews},
  volume={39},
  pages={1024--1034},
  year={2014},
  publisher={Elsevier}
}

@inproceedings{von1999misuses,
  title={Misuses of statistical analysis in climate research},
  author={Von Storch, Hans},
  booktitle={Analysis of Climate Variability: Applications of Statistical Techniques Proceedings of an Autumn School Organized by the Commission of the European Community on Elba from October 30 to November 6, 1993},
  pages={11--26},
  year={1999},
  organization={Springer}
}

@article{katz2010statistics,
  title={Statistics of extremes in climate change},
  author={Katz, Richard W},
  journal={Climatic change},
  volume={100},
  number={1},
  pages={71--76},
  year={2010},
  publisher={Springer}
}

@article{sobrino2004land,
  title={Land surface temperature retrieval from LANDSAT TM 5},
  author={Sobrino, Jos{\'e} A and Jim{\'e}nez-Mu{\~n}oz, Juan C and Paolini, Leonardo},
  journal={Remote Sensing of environment},
  volume={90},
  number={4},
  pages={434--440},
  year={2004},
  publisher={Elsevier}
}

@article{qin2001mono,
  title={A mono-window algorithm for retrieving land surface temperature from Landsat TM data and its application to the Israel-Egypt border region},
  author={Qin, Zhihao and Karnieli, Arnon and Berliner, Pedro},
  journal={International journal of remote sensing},
  volume={22},
  number={18},
  pages={3719--3746},
  year={2001},
  publisher={Taylor \& Francis}
}

@article{marzouk2021assessment,
  title={Assessment of global warming in Al Buraimi, sultanate of Oman based on statistical analysis of NASA POWER data over 39 years, and testing the reliability of NASA POWER against meteorological measurements},
  author={Marzouk, Osama A},
  journal={Heliyon},
  volume={7},
  number={3},
  pages={e06625},
  year={2021},
  publisher={Elsevier}
}

@article{aboelkhair2019assessment,
  title={Assessment of agroclimatology NASA POWER reanalysis datasets for temperature types and relative humidity at 2 m against ground observations over Egypt},
  author={Aboelkhair, Hassan and Morsy, Mostafa and El Afandi, Gamal},
  journal={Advances in Space Research},
  volume={64},
  number={1},
  pages={129--142},
  year={2019},
  publisher={Elsevier}
}

@article{westberg2013analysis,
  title={An analysis of NASA's MERRA meteorological data to supplement observational data for calculation of climatic design conditions},
  author={Westberg, David J and Stackhouse Jr, Paul W and Hoell, James M and Chandler, William S},
  journal={ASHRAE Transactions},
  volume={119},
  pages={210},
  year={2013},
  publisher={American Society of Heating, Refrigeration and Air Conditioning Engineers, Inc.}
}

@article{grinsztajn2022tree,
  title={Why do tree-based models still outperform deep learning on tabular data?},
  author={Grinsztajn, L{\'e}o and Oyallon, Edouard and Varoquaux, Ga{\"e}l},
  journal={arXiv preprint arXiv:2207.08815},
  year={2022}
}

@article{kabo2016multiyear,
  title={Multiyear rainfall and temperature trends in the Volta river basin and their potential impact on hydropower generation in Ghana},
  author={Kabo-Bah, Amos T and Diji, Chuks J and Nokoe, Kaku and Mulugetta, Yacob and Obeng-Ofori, Daniel and Akpoti, Komlavi},
  journal={Climate},
  volume={4},
  number={4},
  pages={49},
  year={2016},
  publisher={MDPI}
}

@inproceedings{randles2017using,
  title={Using the Jupyter notebook as a tool for open science: An empirical study},
  author={Randles, Bernadette M and Pasquetto, Irene V and Golshan, Milena S and Borgman, Christine L},
  booktitle={2017 ACM/IEEE Joint Conference on Digital Libraries (JCDL)},
  pages={1--2},
  year={2017},
  organization={IEEE}
}

@book{kluyver2016jupyter,
  title={Jupyter Notebooks-a publishing format for reproducible computational workflows.},
  author={Kluyver, Thomas and Ragan-Kelley, Benjamin and P{\'e}rez, Fernando and Granger, Brian E and Bussonnier, Matthias and Frederic, Jonathan and Kelley, Kyle and Hamrick, Jessica B and Grout, Jason and Corlay, Sylvain and others},
  volume={2016},
  year={2016}
}

@inproceedings{chockalingam2015agriculture,
  title={Agriculture drought assessment and monitoring (ADAMS) software using ESRI ArcMap},
  author={Chockalingam, Jeganathan and Giriraj, Amarnath and Avishek, Kirti and Mondal, Saptarshi},
  booktitle={16th ESRI India User Conference, Delhi},
  volume={2},
  year={2015}
}

@book{heumann2016introduction,
  title={Introduction to statistics and data analysis},
  author={Heumann, Christian and Shalabh, Micheal Schomaker},
  year={2016},
  publisher={Springer}
}

@online{noaa2020,
    author = {{NOAA}},
    title = {{National Centers for Environmental Information, Global Climate Report - Annual 2015}},
    year = {2020},
    url = {www.ncdc.noaa.gov/sotc/global/202007/supplemental/page-1},
    note = {Accessed 12 September 2020}
}

@article{parastatidis2017online,
  title={Online global land surface temperature estimation from Landsat},
  author={Parastatidis, David and Mitraka, Zina and Chrysoulakis, Nektrarios and Abrams, Michael},
  journal={Remote sensing},
  volume={9},
  number={12},
  pages={1208},
  year={2017},
  publisher={MDPI}
}

@article{jimenez2008revision,
  title={Revision of the single-channel algorithm for land surface temperature retrieval from Landsat thermal-infrared data},
  author={Jimenez-Munoz, Juan C and Cristobal, Jordi and Sobrino, Jos{\'e} A and S{\`o}ria, Guillem and Ninyerola, Miquel and Pons, Xavier},
  journal={IEEE Transactions on geoscience and remote sensing},
  volume={47},
  number={1},
  pages={339--349},
  year={2008},
  publisher={IEEE}
}

@article{jimenez2014land,
  title={Land surface temperature retrieval methods from Landsat-8 thermal infrared sensor data},
  author={Jim{\'e}nez-Mu{\~n}oz, Juan C and Sobrino, Jos{\'e} A and Skokovi{\'c}, Dra{\v{z}}en and Mattar, Cristian and Cristobal, Jordi},
  journal={IEEE Geoscience and remote sensing letters},
  volume={11},
  number={10},
  pages={1840--1843},
  year={2014},
  publisher={IEEE}
}

@book{chambers2015advanced,
  title={Advanced analytics methodologies: Driving business value with analytics},
  author={Chambers, Michele and Dinsmore, Thomas W},
  year={2015},
  publisher={Pearson Education}
}

@article{friedman2001greedy,
  title={Greedy function approximation: a gradient boosting machine},
  author={Friedman, Jerome H},
  journal={Annals of statistics},
  pages={1189--1232},
  year={2001},
  publisher={JSTOR}
}

@article{friedman2002stochastic,
  title={Stochastic gradient boosting},
  author={Friedman, Jerome H},
  journal={Computational statistics \& data analysis},
  volume={38},
  number={4},
  pages={367--378},
  year={2002},
  publisher={Elsevier}
}

@book{kuhn2013applied,
  title={Applied predictive modeling},
  author={Kuhn, Max and Johnson, Kjell and others},
  volume={26},
  year={2013},
  publisher={Springer}
}

@misc{chen2022xgboost,
  title={Xgboost: extreme gradient boosting. R package version 1.6. 0.1},
  author={Chen, Tianqi and He, Tong and Benesty, Michael and Khotilovich, Vadim and Tang, Yuan and Cho, Hyunsu and Chen, Kailong and Mitchell, Rory and Cano, Ignacio and Zhou, Tianyi and others},
  year={2022}
}

@article{monteiro2018assessment,
  title={Assessment of NASA/POWER satellite-based weather system for Brazilian conditions and its impact on sugarcane yield simulation},
  author={Monteiro, Leonardo A and Sentelhas, Paulo C and Pedra, George U},
  journal={International Journal of Climatology},
  volume={38},
  number={3},
  pages={1571--1581},
  year={2018},
  publisher={Wiley Online Library}
}

@article{bessah2022climatic,
  title={Climatic zoning of Ghana using selected meteorological variables for the period 1976--2018},
  author={Bessah, Enoch and Amponsah, William and Ansah, Samuel Owusu and Afrifa, Andrews and Yahaya, Bashiru and Wemegah, Cosmos Senyo and Tanu, Michael and Amekudzi, Leonard K and Agyare, Wilson Agyei},
  journal={Meteorological Applications},
  volume={29},
  number={1},
  pages={e2049},
  year={2022},
  publisher={Wiley Online Library}
}
\newpage
\end{document}